\documentclass[twocolumn]{article}
\usepackage{fancyhdr}
\usepackage{booktabs}
\usepackage{graphicx}
\usepackage{xspace}
\usepackage{multirow}
\usepackage{amssymb,amsmath}
\usepackage{mathtools}
\usepackage{subcaption}
\usepackage{tabularx}
\usepackage{geometry}
\usepackage{fancyvrb}
\usepackage{hyperref}
\usepackage{xcolor}
\usepackage{relsize}
\usepackage{stfloats}
\usepackage{lmodern}
\usepackage[small,compact]{titlesec}
\usepackage{algorithm}
\usepackage{algpseudocode}
\usepackage{cleveref}
\usepackage{cite}
\usepackage{listings}
\usepackage{tikz}
\usepackage[inline]{enumitem}
\usepackage{etoolbox}
\usepackage[most]{tcolorbox}
\usepackage{menukeys}
\usepackage{float}

\makeatletter
\patchcmd{\ttlh@hang}{\parindent\z@}{\parindent\z@\leavevmode}{}{}
\patchcmd{\ttlh@hang}{\noindent}{}{}{}
\makeatother

\newcommand{\codenum}{\ttfamily\tiny\color{gray}}

\tcbset{bbox/.style={breakable,colback=white,colframe=black,
    boxrule=0.4pt,boxsep=0pt,left=2pt,right=4pt}}
\tcbset{nbbox/.style={colback=white,colframe=black,
    boxrule=0.4pt,boxsep=0pt,left=2pt,right=4pt}}
\BeforeBeginEnvironment{algorithmic}{\begin{tcolorbox}[nbbox]}
\AfterEndEnvironment{algorithmic}{\end{tcolorbox}}
\BeforeBeginEnvironment{lstlisting}{\begin{tcolorbox}[bbox]}
\AfterEndEnvironment{lstlisting}{\end{tcolorbox}}
\NewDocumentCommand{\code}{O{}O{bbox}m}{\begin{tcolorbox}[#2]
    \vspace{-10pt}\lstinputlisting[#1]{#3}\vspace{-10pt}\end{tcolorbox}}

\floatstyle{plaintop}
\restylefloat{algorithm}

\usetikzlibrary{positioning, fit, calc, backgrounds, arrows.meta}

\hypersetup{colorlinks, linkcolor={red!50!black}, citecolor={blue!50!black},
  urlcolor={blue!80!black}}

\DeclareRobustCommand{\ccite}[1]{\IfSubStr{#1}{,}{refs.~}{ref.~}\cite{#1}}
\DeclareRobustCommand{\Ccite}[1]{\IfSubStr{#1}{,}{Refs.~}{Ref.~}\cite{#1}}

\newcommand{\fastjet}{F\protect\scalebox{0.8}{AST}%
  J\protect\scalebox{0.8}{ET}\xspace}
\newcommand{\madgraph}{M\protect\scalebox{0.8}{AD}%
  G\protect\scalebox{0.8}{RAPH}\xspace}
\newcommand{\mlhad}{ML\protect\scalebox{0.8}{HAD}\xspace}
\newcommand{\sherpa}{S\protect\scalebox{0.8}{HERPA}\xspace}
\newcommand{\pythia}{P\protect\scalebox{0.8}{YTHIA}\xspace}
\newcommand{\herwig}{H\protect\scalebox{0.8}{ERWIG}\xspace}

\newcommand{\phoenix}{P\protect\scalebox{0.8}{HOENIX}\xspace}
\newcommand{\vistas}{V\protect\scalebox{0.8}{ISTAS}\xspace}
\newcommand{\jupyter}{J\protect\scalebox{0.8}{UPYTER}\xspace}
\newcommand{\colab}{C\protect\scalebox{0.8}{OLAB}\xspace}
\newcommand{\python}{\texttt{Python}\xspace}
\newcommand{\cpp}{\texttt{C++}\xspace}
\newcommand{\json}{\texttt{JSON}\xspace}

\newcommand{\eg}{\textit{e.g.},~}
\newcommand{\ie}{\textit{i.e.},~}
\newcommand{\etc}{\textit{etc.}}

\newcommand{\GeV}{\ensuremath{\text{Ge\kern -0.1em V}}\xspace}
\newcommand{\TeV}{\ensuremath{\text{Te\kern -0.1em V}}\xspace}
\newcommand{\mmp}[1]{~#1}

\newcommand{\display}[2][\textwidth]{%
  \includegraphics[trim={1300px 300px 800px 400px}, clip, width=#1]{#2}}
\newcommand{\event}[4][vistas.txt]{\BVerbatimInput[firstline=#2, lastline=#3,
    formatcom=\color{#4}]{figs/#1}\\[0.5mm]}
\newcommand{\ei}[2]{#1\!\texttt{[#2]}}

\definecolor{hard}{HTML}{D200D4}
\definecolor{MPI}{HTML}{CB0000}
\definecolor{ISR}{HTML}{CD630C}
\definecolor{FSR}{HTML}{B3CE00}
\definecolor{parton}{HTML}{64FF00}
\definecolor{hadronization}{HTML}{0DCE00}
\definecolor{decay}{HTML}{08D27B}
\definecolor{beam}{HTML}{2C00D1}
\definecolor{flow}{HTML}{999999}
\definecolor{other}{HTML}{F2F2F2}
\definecolor{comments}{rgb}{0.0, 0.4, 0.0}

\begin{document}

\twocolumn
\begin{table*}[htb]

\vspace{-7mm}
\rightline{\textsf{FERMILAB-PUB-26-0384-CSAID, MCNET-26-15}}
\vspace{4mm}
  
\begin{center}
  \textbf{\Large \vistas: A Visualization Interface for Particle Collision Simulations}
\end{center}

\begin{center}
Beno\^ it Assi\textsuperscript{1$\spadesuit$},
Christian Bierlich\textsuperscript{2$\clubsuit$},
Rikab Gambhir\textsuperscript{1$\diamond$},
Phil Ilten\textsuperscript{1$\dagger$},
Tony Menzo\textsuperscript{3,4$\star$},
Stephen Mrenna\textsuperscript{1,5$\maltese$},
Manuel Szewc\textsuperscript{6$\parallel$},
Michael K. Wilkinson\textsuperscript{7$\perp$},
Ahmed Youssef\textsuperscript{1$\ddagger$}, 
and Jure Zupan\textsuperscript{1$\mathsection$}
\end{center}

\begin{center}
\textsuperscript{\bf 1} Department of Physics, University of Cincinnati, Cincinnati, Ohio 45221,USA \\
\textsuperscript{\bf 2} Department of Physics, Lund University, Box 118, SE-221 00 Lund, Sweden \\
\textsuperscript{\bf 3} Department of Physics, University of Alabama, Tuscaloosa, AL 35487, USA \\
\textsuperscript{\bf 4} Theory Division, Fermilab, Batavia, Illinois, USA \\
\textsuperscript{\bf 5} Data Science, Simulation and Learning Division, Fermilab, Batavia, Illinois 60510, USA \\
\textbf{\textsuperscript{6}} International Center for Advanced Studies (ICAS), ICIFI and ECyT-UNSAM, 25 de Mayo y Francia, (1650) San Mart\'{i}n, Buenos Aires, Argentina \\
\textsuperscript{\bf 7} Homer L. Dodge Department of Physics and Astronomy,
University of Oklahoma,
Norman, OK 73019, USA\\
${}^\spadesuit${\small \sf assibt@ucmail.uc.edu},
${}^\clubsuit${\small \sf christian.bierlich@hep.lu.se},
${}^\diamond${\small \sf gambhirb@ucmail.uc.edu},
${}^\dagger${\small \sf philten@cern.ch},
${}^\star${\small \sf 
amenzo@ua.edu},
${}^\maltese${\small \sf mrenna@fnal.gov},
${}^\parallel${\small \sf mszewc@unsam.edu.ar},
${}^\perp${\small \sf michael.k.wilkinson@ou.edu},
${}^\ddagger${\small \sf youssead@ucmail.uc.edu},
${}^\mathsection${\small \sf zupanje@ucmail.uc.edu}
\end{center}

\begin{center}
\includegraphics[width=0.4\textwidth]{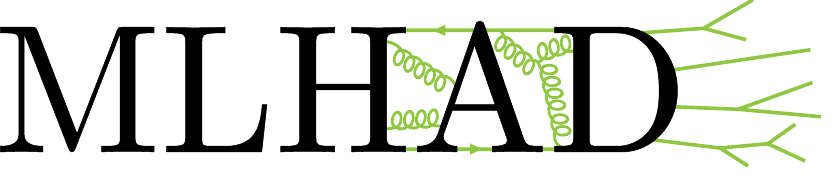}
\end{center}

\begin{abstract}
We introduce \vistas, a tool for visualizing high-energy particle physics collisions simulated by the \pythia Monte-Carlo event generator.
\vistas utilizes the browser-based event display framework \phoenix to show distinct computational stages of a high-energy collision event simulation: the hard process, parton shower, hadronization, and particle decays.
Particles produced from each of these stages are represented as lines in an interactive three-dimensional graph structure, where each line is along the direction of its particle's three-momentum vector.
The event can be rotated, translated and zoomed, and details for each particle can be accessed by selecting the relevant particle line.
Additionally, particle lines from all stages of the simulation can be toggled on and off and can be filtered by particle-level kinematic selection requirements.
This interactive environment provides an intuitive interpretation of \pythia simulation output, including detailed features such as color flow, beam remnants, and multiple parton interactions, making it a useful tool in physics education settings, from outreach activities to graduate particle-physics courses.
\end{abstract}
\end{table*}

In high-energy particle physics (HEP), accelerators are used to collide particles such as protons, electrons, and positrons.
In each collision, called an \textit{event}, particles are produced and are reconstructed by complex detectors, with the goal to infer the properties of fundamental physics and particle interactions.
Crucially, theoretical predictions are needed to perform such inference and connect measurements with models of nature.
Many of these predictions are made with Monte-Carlo event generators (MCEGs)\cite{Buckley:2011ms,Campbell:2022qmc}, where pseudo-random number sequences are combined with numerical techniques to perform calculations based on the underlying quantum field theory (QFT).
MCEGs provide theoretical predictions for the event rates and simulate representative samples of collision events.
In these simulated events MCEGs produce, with probabilities that follow theoretical expectations, final-state particles that are then observed in detectors.

To further understand the operation of HEP colliders, visualization techniques for final-state particles in detectors, such as the \textsc{Root}-based event displays\cite{Brun:1997pa}, ATLAS's VP1 and \textsc{Atlantis}\cite{Konstantinidis:2005zz,Atlantis}, ALICE's \textsc{AliEve} and \textsc{RenderCore}\cite{Niedziela:2017oad,Bohak:2023fij}, CMS's \textsc{iSpy}\cite{iSpy}, and LHCb's \textsc{LHCb Event Display}\cite{Trisovic:2014kir}, have been introduced.\footnote{
ATLAS, ALICE, LHCb, and CMS are four large detectors on the Large Hadron Collider (LHC), based at the European Organization for Nuclear Research (CERN) located in Geneva, Switzerland.}
In HEP education, visualizing the final-state particles in detectors gives a schematic representation of how the detector reconstructs different types of particles, which helps students to understand what a particle-physics event looks like in a detector.
Indeed, the utility of particle-physics event visualization goes well beyond education, and is also used by experts in the day-to-day operation of large HEP detectors\cite{Bianchi:2017dzy,Kittelmann:2010zz,VP1,Alverson:2012gd,iSpy}.
Most of the current event displays focus exclusively on visualizing only the final-state particles and how they interact with the detector, and these visualization techniques are quite mature.
In contrast, however, to these final-state detector visualizations, visualization of the MCEG computational pipeline that predicts those particles remains comparatively underdeveloped.\footnote{
For instance, \madgraph\cite{Alwall:2014hca} computes the \textit{hard process} using perturbative QFT methods and can render the corresponding Feynman diagrams\cite{Feynman:1949zx}, but provides no visualization of the remaining stages of the event generation.}

\begin{figure}
  \begin{center}
    \includegraphics[width=\columnwidth]{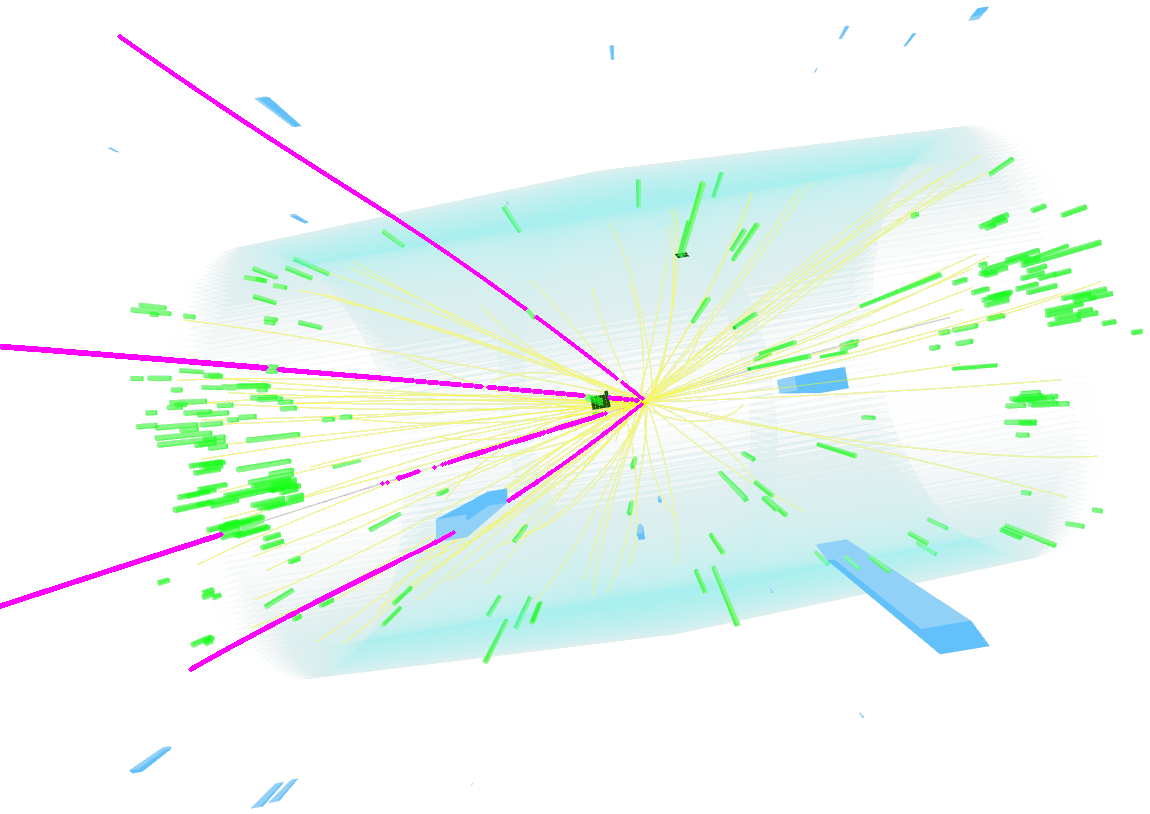}
  \end{center}
  \caption{\label{fig:cms}Example event reconstructed by the CMS detector consistent with a Higgs boson decaying into two muons and two anti-muons, represented by magenta lines.
    This image was produced using the \textsc{iSpy}\cite{Alverson:2012gd,iSpy} visualization tool and CMS open data.
    Yellow lines represent reconstructed trajectories for charged particles and the blue and green rectangles represent measured energy deposits.}
\end{figure}

\begin{figure}
  \begin{center}
    \includegraphics[width=\columnwidth]{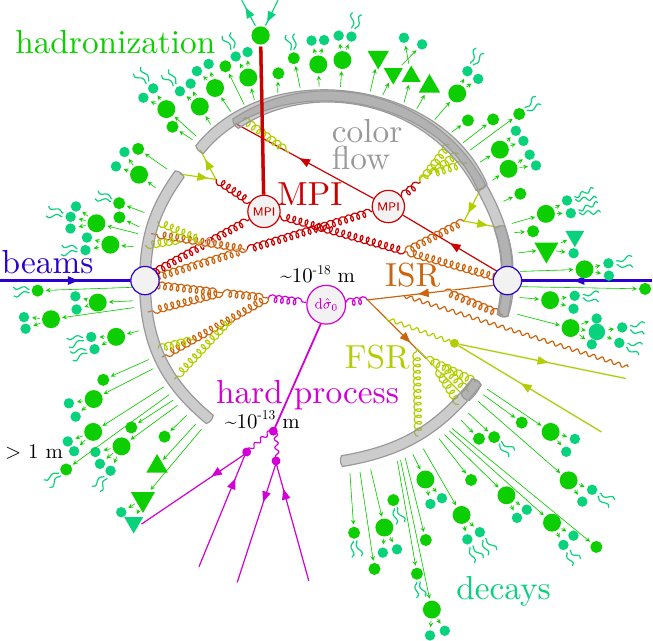}
  \end{center}
  \caption{\label{fig:pythia}Two dimensional graphical representation of a sample HEP collision event simulated by the \pythia MCEG for a Higgs boson decaying into two muons and two anti-muons.
    The final-state particles are denoted as the green and cyan lines, circles, or triangles at the outer edges of the figure and the blue lines are the incoming proton beams.
    This image was modified from the \pythia manual\cite{Bierlich:2022pfr}.
    \vistas generalizes to three dimensions this type of graphical representation of HEP collision events.}
\end{figure}

In this paper, we introduce the \textit{V}isualization \textit{I}nterface for \textit{S}imulated \textit{T}opologies and \textit{A}nalysis of \textit{S}cattering (\vistas), a new educational tool for interactively exploring the full event record produced by the \pythia MCEG\cite{Bierlich:2022pfr}.
Using the HEP Software Foundation (HSF) browser-based \phoenix event viewer\cite{Moyse:2021xai,Phoenix}, \vistas displays the entire MCEG computational pipeline as a fully interactive three-dimensional render, similar to existing detector-level tools, allowing for the visualization and analysis of the theoretical structure of particle collisions.

With \vistas, we address a major challenge of visualizing full MCEG simulations: large scale separations, where physical processes within a single event can span over 18 orders of magnitude in length.
As an example, consider a collision in which a Higgs boson is produced and subsequently decays into two muons and two anti-muons, which, together with other final-state particles, can be reconstructed by detectors such as ATLAS\cite{ATLAS:2008xda} and CMS\cite{CMS:2008xjf}.
The Higgs boson is produced at a length scale of $10^{-18}$ meters and decays less than $10^{-13}$ meters away from the collision point, yet the final-state particle decays and parts of the physical detector apparatus are located at scales of millimeters to meters away.

\begin{figure}
  \begin{center}
    \display[0.47\textwidth]{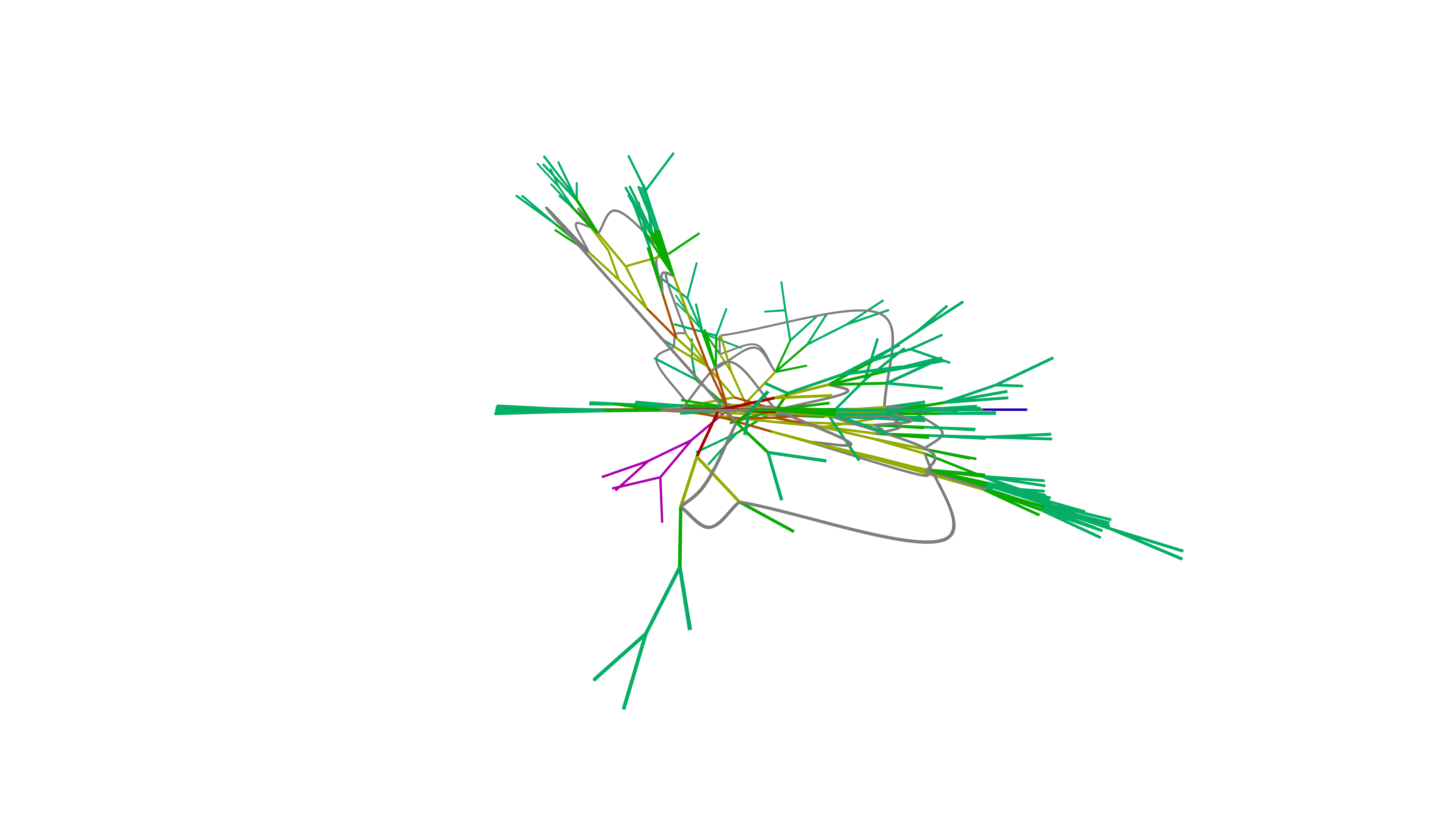}
  \end{center}
  \caption{\label{fig:vistas}Screen capture of a sample HEP collision event simulated by the \pythia MCEG for a Higgs boson decaying into two muons and two anti-muons visualized with \vistas.
    The same color scheme as \cref{fig:pythia} is used.
    An interactive version of this plot is available at \url{https://cern.ch/vistas-paper} and the code used to produce this plot is given in \cref{app:code}.}
\end{figure}

In \cref{fig:cms,fig:pythia,fig:vistas} we illustrate three different ways of visualizing this collision:
\begin{enumerate*}[label=(\arabic*)]
\item a detector-level rendering of final-state particles in \cref{fig:cms},
\item a two-dimensional graphic with approximate representations of the intermediate physics in \cref{fig:pythia}, and
\item a full three-dimensional representation of the theoretical intermediate states with \vistas in \cref{fig:vistas}.
\end{enumerate*}
These representations serve distinct purposes and vary in several meaningful ways.
Because the Higgs boson is unstable and decays well before reaching any detector element, it is not a final-state particle and does not appear in \cref{fig:cms}.
It is typical in HEP events for interesting intermediate particles, \eg the Higgs boson, to not directly leave any detector signal, and instead their properties must be inferred.
Only particles that travel distances of millimeters to meters before decaying can interact with the detector, such as charged pions or muons.

The two-dimensional graphic in \cref{fig:pythia} shows both final-state and intermediate particles, with a focus on the Higgs decay.
The disparate length scales are visualized by not requiring a one-to-one length mapping in the diagram.
The diagram zooms in on short-distance processes like the Higgs boson production and decay in magenta, and zooms out on long-distance processes like stable particles.
Everything within the gray circular bands of \cref{fig:pythia} would appear in the detector-level rendering of \cref{fig:cms} as a single point, indistinguishable from the collision point, but \cref{fig:pythia} is still a two-dimensional schematic, not a full three-dimensional rendering of a specific event.
The \vistas representation in \cref{fig:vistas} combines the advantages of \cref{fig:pythia} and \cref{fig:cms}, focusing on the internal structure of the event while also retaining key features of existing detector-level tools such as rendering specific events and keeping an interactive display.

This paper is organized as follows:
\Cref{sec:pythia} contains a brief overview of the \pythia simulation process, the resulting event record, and how this connects to event displays from \vistas.
The architecture of \vistas is described in \cref{sec:vistas}, \vistas usage is described in \cref{sec:usage}, and conclusions are given in \cref{sec:conclusions}.

\section[Pythia simulation review]{\pythia simulation review}
\label{sec:pythia}

The major MCEGs in HEP (\pythia\cite{Bierlich:2022pfr}, \herwig\cite{Bahr:2008pv}, and \sherpa\cite{Sherpa:2024mfk}) all have slightly different methods of simulating events and internally representing the event record.
In \vistas we have focused on visualizing events as produced by \pythia, although the code could be modified to accommodate other MCEGs.
The simulation of high-energy particle collisions via \pythia\cite{Sjostrand:2006za,Sjostrand:2014zea,Bierlich:2022pfr} proceeds through several well-defined stages, producing an event record that is a list of particles.
We describe this event record in \cref{sec:pythia:event} and then describe the stages used to create this event record in \cref{sec:pythia:stages}.

\subsection{The event record}
\label{sec:pythia:event}

\begin{table*}
  \caption{\label{tab:event}Abridged version of the \pythia event record used to produce the visualized event of \cref{fig:vistas,fig:stages}.
    Each row corresponds to a particle, and each column information for that particle as described in \cref{sec:pythia:event}.
    The color coding corresponds to the simulation stages of \pythia described in \cref{sec:pythia:stages} and matches the color scheme of \cref{fig:vistas}: \textcolor{hard}{\textit{hard process} (magenta)}, \textcolor{beam}{\textit{beams} (blue)}, \textcolor{MPI}{\textit{MPI} (red)}, \textcolor{ISR}{\textit{ISR} (orange)}, \textcolor{FSR}{\textit{FSR} (yellow)}, \textcolor{hadronization}{\textit{hadronization} (green)}, \textcolor{decay}{\textit{decay} (cyan)}.
    The record has been truncated to show a smaller number of representative rows; there are 633 particles total in the full record.}
  \relscale{0.65}
  \begin{center}
    \event{0}{5}{black}
    \event{6}{7}{beam}
    \event{8}{10}{hard}
    \event{11}{14}{ISR}
    \texttt{...} \\[2mm]
    \event{39}{39}{FSR!85!black}
    \texttt{...} \\[2mm]
    \event{43}{44}{FSR!85!black}
    \texttt{...} \\[2mm]
    \event{92}{95}{MPI}
    \texttt{...} \\[2mm]
    \event{262}{267}{hard}
    \texttt{...} \\[2mm]
    \event{280}{283}{hadronization}
    \event{284}{289}{hadronization}
    \texttt{...} \\[2mm]
    \event{436}{437}{decay}
    \texttt{...} \\[2mm]
    \event{568}{569}{decay}
    \texttt{...} \\[2mm]
    \event{638}{638}{decay}
    \event{639}{641}{black}
  \end{center}
\end{table*}

A comprehensive description of the \pythia event record can be found in the \pythia physics manual\cite{Bierlich:2022pfr}; a summary is provided here.
An abridged version of the \pythia event record used to create \cref{fig:vistas,fig:stages} is given in \cref{tab:event}.
Each row in the event record corresponds to a particle produced in the simulation and each column is a particle property.
\begin{itemize}
\item \texttt{no}: the index of the particle in the event record.
\item \texttt{id}: the Particle Data Group\cite{ParticleDataGroup:2024cfk} identification number (PDG ID) for the type of the particle. This is a common way of encoding the particle type in MCEGs.
\item \texttt{name}: a more human readable version of the particle type.
  The particle with index \texttt{1} has a PDG ID of \texttt{2212} and a name of \texttt{p+}.
  This particle is a proton and one of the incoming LHC beams in this event.
  A name in parentheses indicates an intermediate particle.
\item \texttt{status}: the \pythia status of the particle, corresponding to the stage of the simulation which produced this particle.
  A negative status indicates an intermediate particle.
  In \cref{tab:event}, we have color-coded entries by status.
\item \texttt{mothers}: range of indices for the particles that produced this particle, commonly referred to as mothers.
  The Higgs boson with index \texttt{5} has two mothers with indices \texttt{3} and \texttt{4} which correspond to gluons coming from the proton beams.
  There are some subtleties of how this index range is interpreted that we omit here for clarity.
\item \texttt{daughters}: range of indices for the particles produced from this particle, commonly referred to as daughters.
  The Higgs boson with index \texttt{5} has one daughter with index \texttt{8} which is another instance of the same Higgs boson, but with modified kinematics due to the emission of additional less energetic particles in the event.
\item \texttt{colours}: the quantum chromodynamics (QCD) charge carried by the particle.
  In most MCEGs, the large color limit approximation\cite{tHooft:1973alw} is used to represent QCD charge.
  In this limit, gluons carry two color indices, a color and an anti-color. 
  In the event record, the first number is the color of the particle and the second number is the anti-color.
  The incoming gluon with index \texttt{3} in \cref{tab:event} carries color \texttt{101} and anti-color \texttt{102}.
  The other incoming gluon with index \texttt{4} carries color \texttt{102} and anti-color \texttt{101}.
  Together, these two gluons have no net color.
  In \pythia, colors typically start from \texttt{101}.
\item \texttt{p\_x}, \texttt{p\_y}, \texttt{p\_z}, \texttt{e}: the four-momentum $p^\mu$ of the particle corresponding to the $x$, $y$, $z$ components of the three-momentum vector $\vec{p}$, and the energy of the particle.
  Here, the $z$ axis is defined along the beam direction.
  \pythia uses natural units, where the speed of light $c$ is set to one, and the four-momentum is expressed in \GeV.
\item \texttt{m}: mass of the particle.
  This can be calculated from the particle four-momentum but is stored for numerical precision, which is particularly relevant for very boosted particles.
  Note that this is not necessarily the on-shell mass of the particle.
\end{itemize}

Beyond the information listed in \cref{tab:event} the event can optionally be listed with the following additional information.
\begin{itemize}
\item \texttt{scale}: the scale, in units of \GeV, at which the particle was produced in the event. See \cref{sec:pythia:stages} for details on the ordering of this scale.
\item \texttt{pol}: the particle polarization, which has no units.
\item \texttt{xProd}, \texttt{yProd}, \texttt{zProd}: the position of where the particle was produced in units of millimeters. This assumes that the particles in the event are not experiencing any external force, such as that from a magnetic field.
\item \texttt{tau}: the invariant lifetime of the particle in units of millimeters$/c$.
\end{itemize}
The position and lifetime information is primarily intended for particles that may travel a significant distance before decaying.

The \pythia event record can be thought of as a directed graph where particle production and decay instances are the  vertices.\footnote{
Common notation for vertices in a graph is also nodes.
Here, a graph is a mathematical structure used to model pairwise relationships between objects, which are taken to be the nodes of the graph.
The graph structure of the \pythia event record is somewhat hidden.
The particles, which form the edges of the graph, are explicitly listed in the event record, while the vertices need to be determined from the event record.
The \vistas display of the event record makes this graph structure more explicit.
Graph vertices are the points from which particle lines either start or end, see \eg \cref{fig:stages} below.
For a non-technical introduction to graph theory see \ccite{Chartrand:1984}.
The \textsc{HepMC3} format\cite{Buckley:2019xhk} is oftentimes used to store HEP events in a graph structure.}
The vertices are linked by the incoming and outgoing particles.
These act as directed edges of the graph, where the direction of the edge is chronological from an earlier time in the simulation to a later time.
In \cref{tab:event}, an example vertex $\ei{g}{3}\, \ei{g}{4} \to \ei{H}{5}$\footnote{
With the $\ei{X}{i}$ notation we reference the event record index \texttt{i} for particle $X$.}
has incoming particles \texttt{3} and \texttt{4}, two gluons from the proton beams, and an outgoing Higgs boson with index \texttt{5}.
This vertex is labeled as $\text{d}\hat{\sigma}_0$ in \cref{fig:pythia} and is the origin of the magenta tree-like structure in \cref{fig:vistas} near the center of the event.

\subsection{Event stages}
\label{sec:pythia:stages}

\begin{figure*}
  \begin{center}
    \begin{subfigure}{0.47\textwidth}
      \display{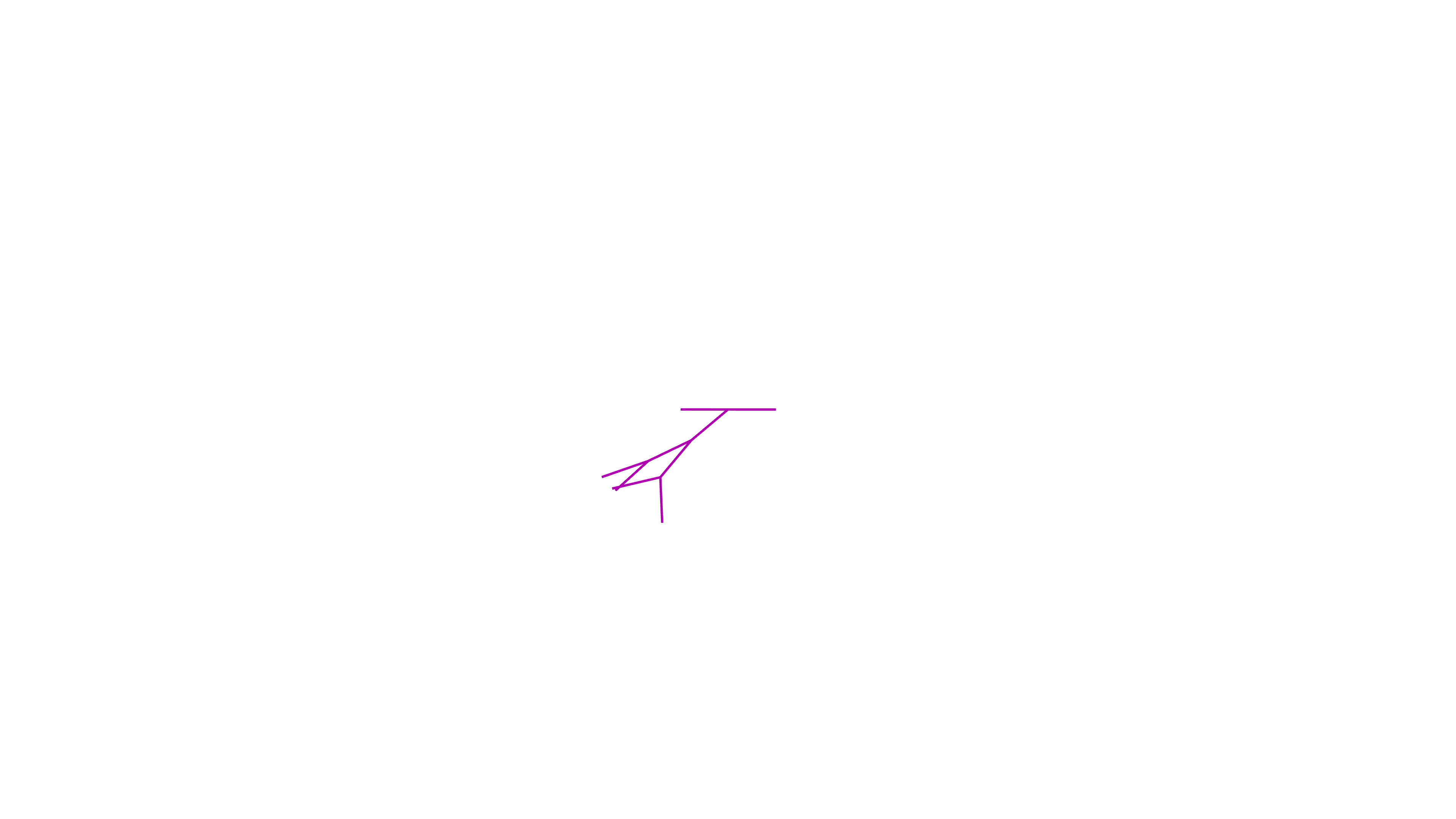}
      \caption{\label{fig:stages:hard}Hard process.}
    \end{subfigure} \hfill
    \begin{subfigure}{0.47\textwidth}
      \display{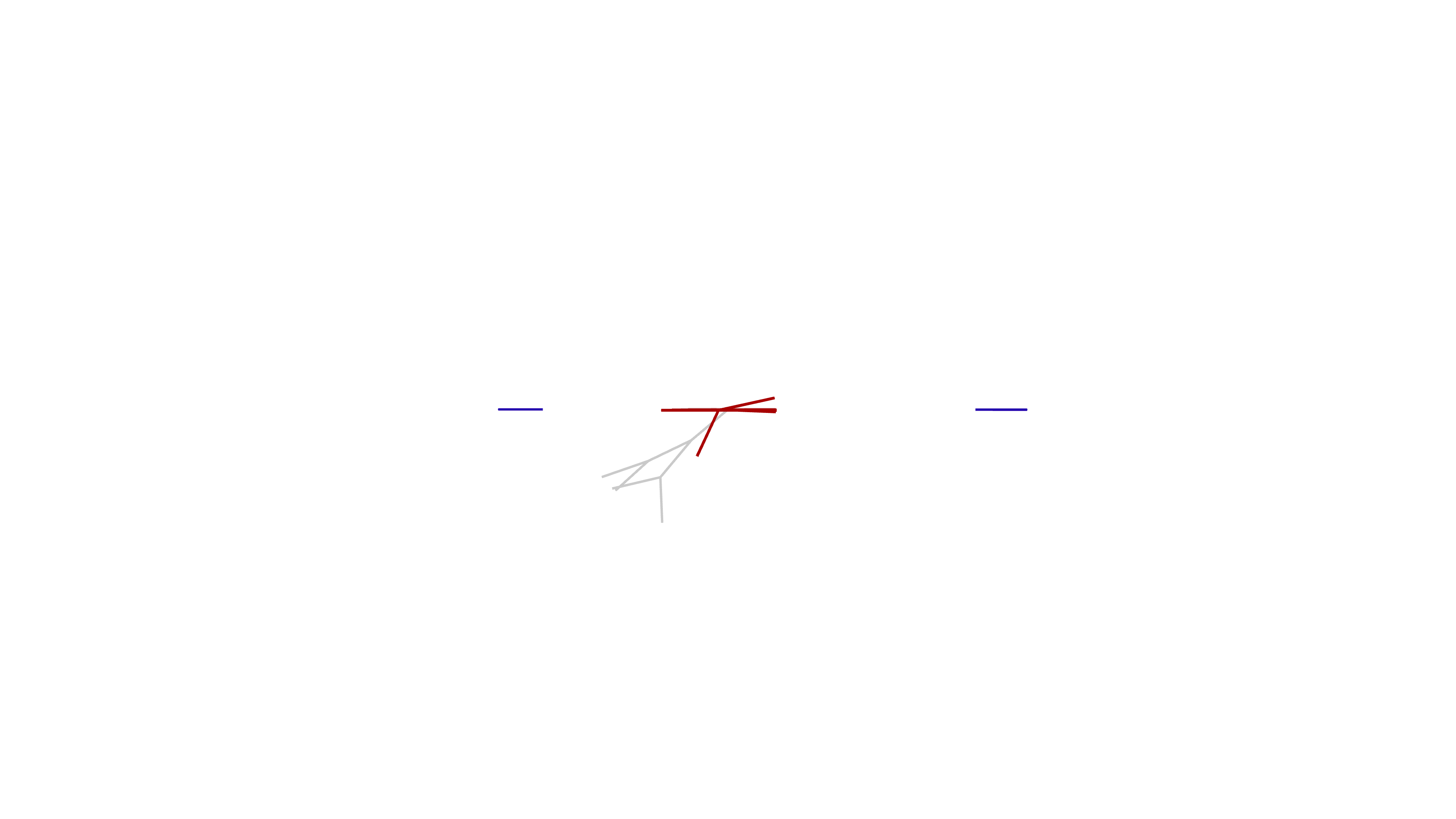}
      \caption{\label{fig:stages:mpi}Multi-parton interactions (red) and beams (blue).}
    \end{subfigure} \\
    \begin{subfigure}{0.47\textwidth}
      \display{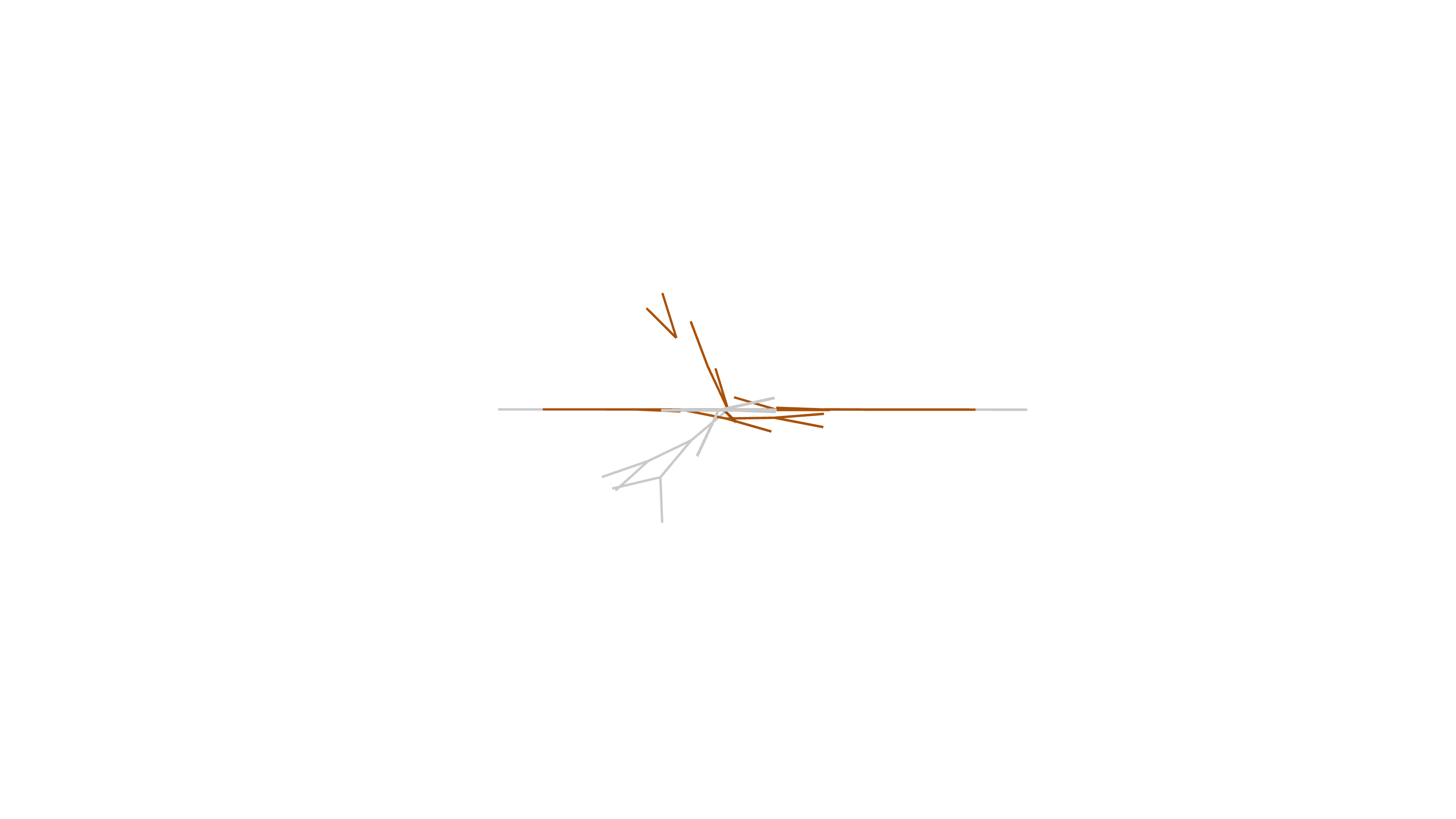}
      \caption{\label{fig:stages:isr}Initial-state radiation.}
    \end{subfigure} \hfill
    \begin{subfigure}{0.47\textwidth}
      \display{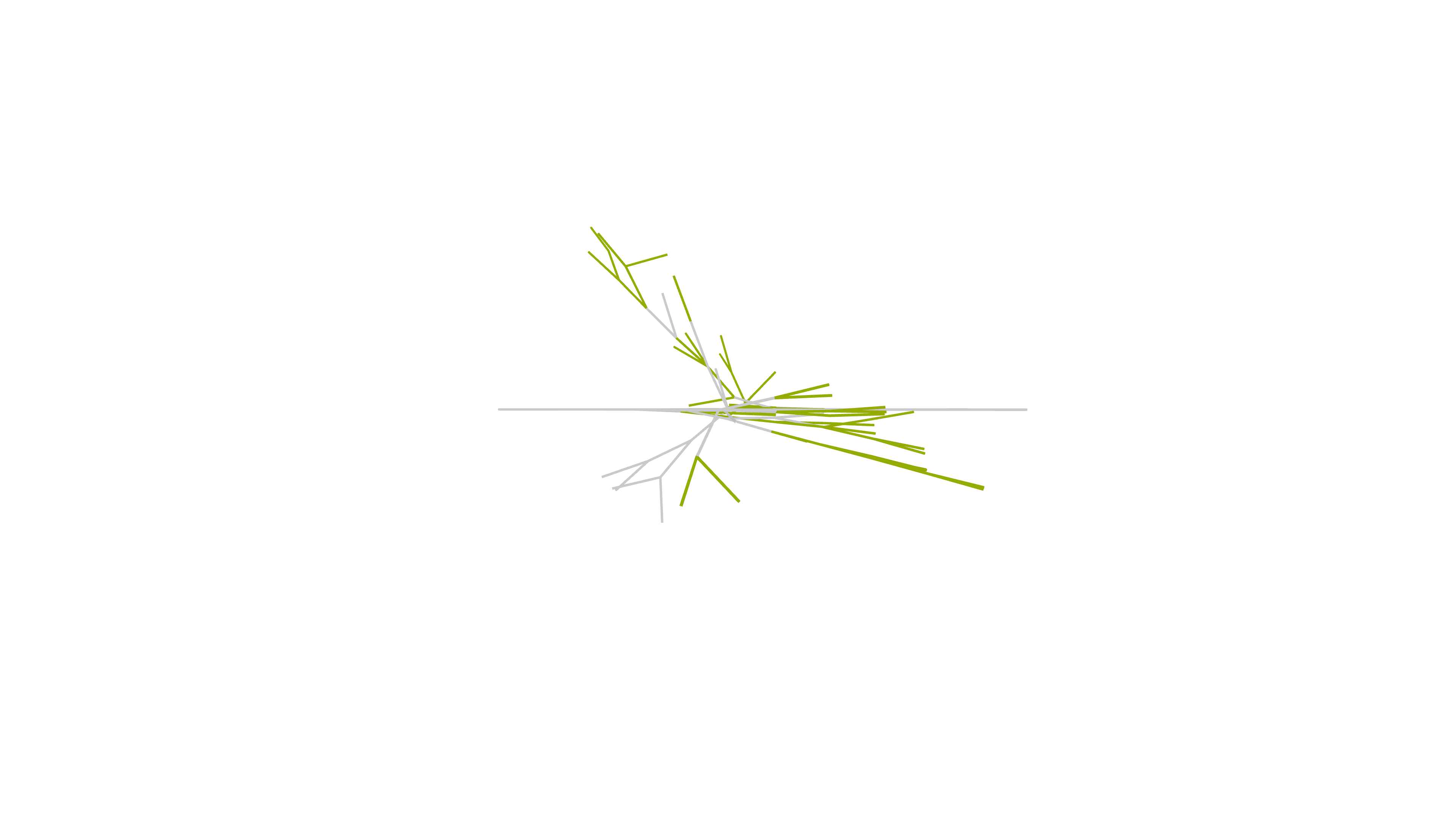}
      \caption{\label{fig:stages:fsr}Final-state radiation.}
    \end{subfigure} \\
    \begin{subfigure}{0.47\textwidth}
      \display{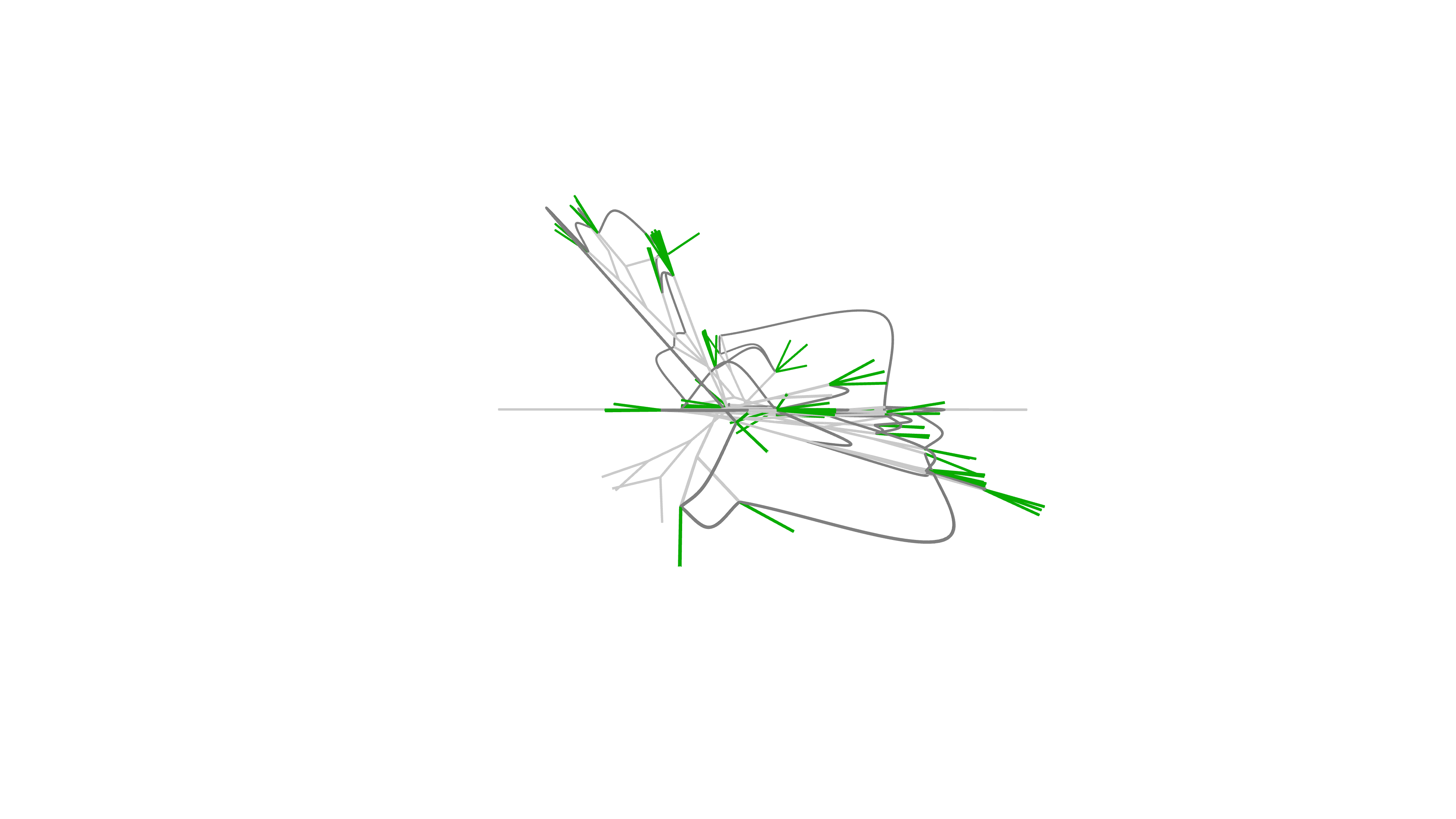}
      \caption{\label{fig:stages:hadronization}Hadronization (green) and strings (dark gray).}
    \end{subfigure} \hfill
    \begin{subfigure}{0.47\textwidth}
      \display{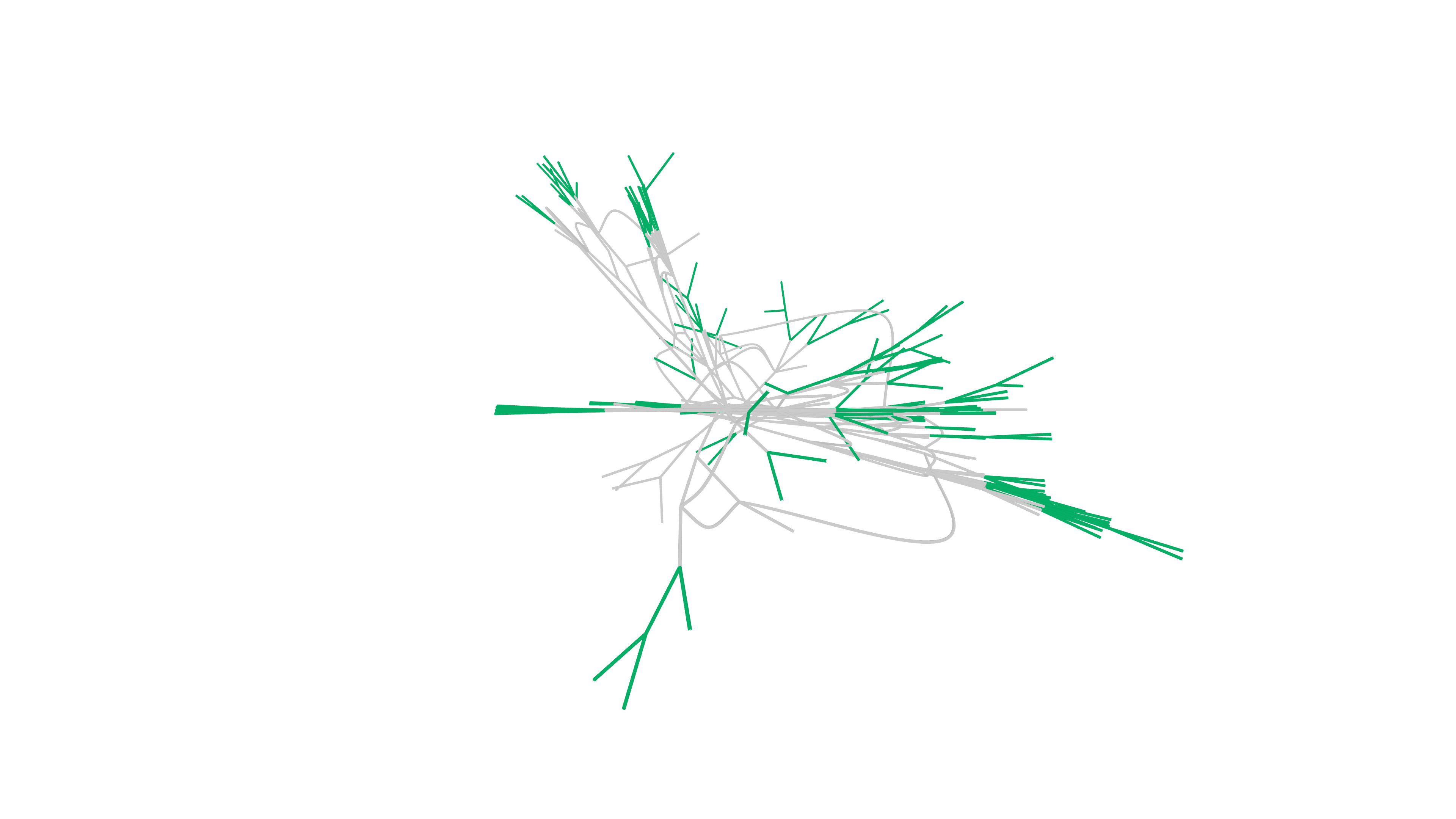}
      \caption{\label{fig:stages:decay}Decay.}
    \end{subfigure}
  \end{center}
  \caption{\label{fig:stages}\vistas visualization of the \pythia simulation stages from the event of \cref{tab:event}.}
\end{figure*}

\pythia simulates events over several stages.
The \texttt{status} codes in \cref{tab:event} indicate during which stage of the simulation the particle was produced.
The simulation sequence is ordered by an energy-like scale, beginning at the highest (hardest) energy scale and evolving downward to the lowest (softest) energy scale.
In this evolution, stages of the simulation can be repeated, or interleaved, at different scales\cite{Sjostrand:2004ef}.
Below we describe each stage of \pythia, keeping in mind that the ordering of the stages is not fixed and can be interleaved.
In this list, the status code range for each \pythia simulation stage is given in the square brackets after the name of the stage.
For clarity, \cref{fig:stages} also shows the particles produced at different simulation stages, visualized using the \vistas event display.

\paragraph[hard process]{\textcolor{hard}{\textit{hard process} [\texttt{21} -- \texttt{29}]:}}
Most events produced at hadron colliders like the LHC are not of particular interest, consisting of relatively low-energy interactions between quarks and gluons.
Instead, we are typically interested in higher-energy processes, like the production of a Higgs boson in the example of \cref{fig:pythia,fig:vistas,tab:event}.
For efficiency and computational tractability, \pythia can be configured to only simulate these processes of interest.
This stage of the event is called the hard process and is the beginning point of the simulation at the hardest scale of the event.\footnote{
While in nature initial-state radiation would be emitted before the hard process, this chronological order is not followed in the \pythia simulation, where processes occurring at the highest energy scale are simulated first.
Initial-state radiation is a process associated with lower energy scales and thus enters the simulation pipeline at a later stage.}
In the event record in \cref{tab:event}, the hard process contains particles \texttt{3}, \texttt{4}, and \texttt{5}, corresponding to the process $\ei{g}{3}\,\ei{g}{4} \to \ei{H}{5}$, and are denoted with magenta lines in \cref{fig:pythia,fig:vistas}.

The hard process is calculated using perturbative QFT (pQFT).
This is usually only applied to fundamental particles, and thus the hard process typically does not contain composite particles such as the initial proton or other hadrons.\footnote{
Hadrons are particles composed of quarks and gluons, held together by strong force (QCD interactions).
Examples include the proton, neutron, pion, kaon, \etc}
The decays of intermediate particles created in the hard process are also typically included as part of the hard process, as long as their decays are calculated using pQFT.
This is visible in \cref{fig:stages:hard}, which visualizes the hard process of the example event; the full event is shown in \cref{fig:vistas}.
It is clear that the hard process consists of more than just Higgs production via gluon fusion,  $g\,g \to H$.
In \cref{fig:stages:hard}, the horizontal line denotes the incoming gluons and the line branching off downwards denotes the Higgs boson.
The Higgs boson then decays into two $Z$ bosons, $H\to ZZ$, where each of these $Z$ bosons subsequently decays into a $\mu^+$ and a $\mu^-$, $Z \to \mu^+\, \mu^-$.
The whole hard process chain is thus $g\, g \to H \to Z\, Z \to (\mu^+\, \mu^-)\, (\mu^+\, \mu^-)$.
The end of this decay tree, $\ei{Z}{257}(\to \ei{\mu^-}{259}\, \ei{\mu^+}{260})\,\ei{Z}{258}(\to \ei{\mu^-}{261}\, \ei{\mu^+}{262})$, is also visible in \cref{tab:event}.

\paragraph[beams]{\textcolor{beam}{\textit{beams} [\texttt{11} -- \texttt{19}]:}}
The hard process must be connected back to the colliding beam particles.
This is done through parton distribution functions\cite{Feynman:1969ej,Bjorken:1969ja,Collins:1985ue} which encode the probability of extracting the incoming particles of the hard process from the beams.
In the example of \cref{tab:event}, the two beams are given by two protons, $\ei{p}{1}$ and $\ei{p}{2}$.
These beams are highlighted in blue in \cref{fig:stages:mpi}.

\paragraph[MPI]{\textcolor{MPI}{\textit{multi-parton interactions (MPI)} [\texttt{21} -- \texttt{29}]:}}
After the hard process extracts particles from the beams, the resulting beam remnants can undergo further interactions.
If the beam particles are hadrons like the proton, other quarks and gluons within these beam remnants may interact.
These additional multi-parton interactions (MPI) are typically QCD processes, such as $g\, g \to g\, g$ or $g\, q \to g\, q$, and, similar to the hard process, are calculated using pQFT.
The incoming particles for the MPI must also be connected back to the beams using parton distribution functions.
The example MPI process $\ei{g}{87}\, \ei{d}{88} \to \ei{g}{89}\, \ei{d}{90}$ in \cref{tab:event} is highlighted in red in \cref{fig:stages:mpi}.

\paragraph[initial-state radiation]{\textcolor{ISR}{\textit{initial-state radiation (ISR)} [\texttt{41} -- \texttt{49}]:}}
When connecting the hard process or MPI back to the beams, the simulated hard process particles may radiate additional particles.
This corresponds to physical processes that occur before the hard process collisions.
For instance, in an electron-positron collision, the incoming $e^-$ or $e^+$ may radiate a photon before the hard collision occurs.
Similarly, gluon and quark systems may radiate additional particles.
The radiation of additional particles can be factorized into a sequential algorithm called a parton shower\cite{Fox:1979ag}, where a tree of emissions is built using pQFT predicted probabilities for each emission.
For initial-state radiation (ISR), this evolution proceeds backwards in time in the MCEG simulation, starting from whatever hard process particles are colliding and evolving back to the initial beam\cite{Sjostrand:1985xi}.

An example ISR from \cref{tab:event} is the process $\ei{u}{6} \to \ei{u}{9}\, \ei{g}{3}$, where the gluon is an incoming hard process particle.
Note that in the index range \texttt{6} -- \texttt{9} there is another Higgs boson, $\ei{H}{8}$.
This is a copy of the hard-process Higgs boson $\ei{H}{5}$ with a shifted four-momentum due to ISR.
In \vistas, particle copies from ISR or other mechanisms are removed by default to simplify the visualization.

The ISR particles listed in \cref{tab:event} are highlighted in \cref{fig:stages:isr}, where this specific ISR example corresponds to the horizontal orange line coming off the hard process to the right.
In \cref{fig:stages:isr} there are missing links in the ISR graph; this is because of the interleaved evolution of \pythia, where the links correspond to interleaved final-state radiation.

\paragraph[final-state radiation]{\textcolor{FSR}{\textit{final-state radiation (FSR)} [\texttt{51} -- \texttt{59}]:}}
Particles may also radiate additional particles forward in time.
This is called final-state radiation (FSR) and is also part of the parton shower which uses pQFT to calculate each emission.
An example FSR from \cref{tab:event} is $\ei{g}{34} \to \ei{g}{38}\, \ei{g}{39}$, where a gluon emits another gluon.
The FSR is highlighted in \cref{fig:stages:fsr} and provides additional emissions while also filling in the missing links of the ISR graph.

\paragraph[hadronization]{\textcolor{hadronization}{\textit{hadronization} [\texttt{71} -- \texttt{89}]:}}
The parton shower, both ISR and FSR, is not valid when the underlying QFT is no longer perturbative.
This occurs when the couplings of the underlying QFT become sufficiently large for the perturbation series to no longer converge.
In the standard model of particle physics, the coupling for QCD rapidly increases at lower energy scales.
Practically, for particles that carry a non-zero QCD charge, pQFT fails at energies around $1~\GeV$ and lower.
Below this energy scale, parton showers are no longer valid and so a different model must be used.

Hadronization is the process by which particles with QCD charge, quarks and gluons, are combined into composite particles with no net QCD charge, hadrons.
There are two major hadronization models, the cluster model\cite{Webber:1983if} and the string model\cite{Andersson:1983ia}; \pythia uses the string model.
In the string model, quarks and gluons are first grouped together into collections of particles with no net QCD charge.
An example of such a collection of particles in \cref{tab:event} are $\ei{u}{275}$, $\ei{g}{276}$, $\ei{g}{277}$, and $\ei{\bar{d}}{278}$. 

We can see how this collection has no net QCD charge in the large-color limit used by \pythia.
The color of $\ei{u}{275}$ is \texttt{106}, which is canceled by the anti-color of $\ei{g}{276}$.
The color \texttt{149} of $\ei{g}{276}$ is canceled by the anti-color of $\ei{g}{277}$.
Finally, the color \texttt{146} of $\ei{g}{277}$ is canceled by the anti-color of $\ei{\bar{d}}{278}$.

These QCD charge cancellations encode the final color flow of the event, and are visualized in \cref{fig:stages:hadronization} by the dark-gray curved lines.
In the string model, a contiguous group of these lines is called a string.
These strings are then fragmented into hadrons like protons or pions.
The hadrons produced from the string example of \cref{tab:event} are particles \texttt{279} -- \texttt{284}.
In \cref{fig:stages:hadronization}, the hadrons produced from the string are given by the lines in green.

\paragraph[decay]{\textcolor{decay}{\textit{decay} [\texttt{91} -- \texttt{99}]:}}
Many of the hadrons produced from hadronization are not stable on the length scale of the detector and will decay before interacting with it.
For example, the $\rho^-$ lifetime is approximately $5\times10^{-24}$ seconds\cite{ParticleDataGroup:2024cfk} and decays primarily into a charged and neutral pion, $\pi^-$ and $\pi^0$.
In \cref{tab:event}, the process $\ei{\rho^-}{284} \to \ei{\pi^-}{431}\, \ei{\pi^0}{432}$ illustrates this type of decay, where the $\ei{\rho^-}{284}$ was created during the hadronization stage.
\pythia continues to decay particles until stable final-state particles are reached.
Here, the $\ei{\pi^-}{431}$ is stable, but the $\ei{\pi^0}{432}$ is not, which subsequently decays into two photons, $\ei{\pi^0}{432} \to \ei{\gamma}{563}\, \ei{\gamma}{564}$.
The decays for the example event in \cref{tab:event} are visualized in \cref{fig:stages:decay}.
Decays are calculated in \pythia using a variety of methods ranging from pQFT to phenomenological models.

\section[Vistas architecture]{\vistas architecture}
\label{sec:vistas}

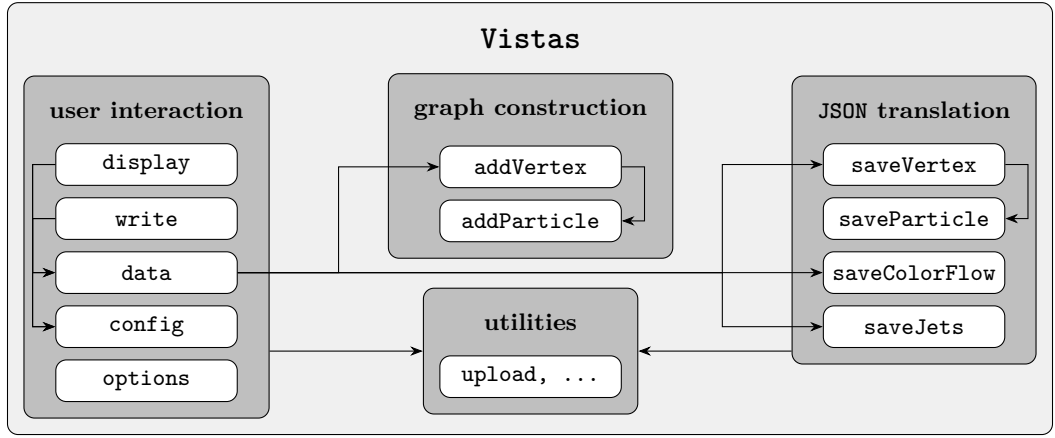
\begin{figure*}
  \vspace{-1cm}
  \begin{center}
    \pgfdeclarelayer{outer}
\pgfsetlayers{outer,background,main}
\newlength{\msep}
\setlength{\msep}{0.15cm}
\newlength{\hsep}
\setlength{\hsep}{6pt}
\begin{tikzpicture}[
  elem/.style={draw, rounded corners, fill=white,
               minimum width=2.4cm, minimum height=0.55cm,
               font=\ttfamily\small},
  group/.style={draw, rounded corners, fill=gray!50, inner sep=6pt},
  outer/.style={draw, thin, rounded corners, fill=gray!12, inner sep=6pt},
  title/.style={font=\bfseries\small},
  bigtitle/.style={font=\bfseries\large},
]

\node[title] (ui) {user interaction};
\node[elem, below=\hsep of ui] (de) {display};
\node[elem, below=\msep of de] (we) {write};
\node[elem, below=\msep of we] (dat) {data};
\node[elem, below=\msep of dat] (cfg) {config};
\node[elem, below=\msep of cfg] (opt) {options};
\begin{pgfonlayer}{background}
  \node[group, fit=(ui)(de)(we)(dat)(cfg)(opt)] (uig) {};
\end{pgfonlayer}
\draw[-Stealth] (de.west) -- ++(-3mm,0) |- (dat.west);
\draw[-Stealth] (de.west) -- ++(-3mm,0) |- (cfg.west);
\draw[-Stealth] (we.west) -- ++(-3mm,0) |- (dat.west);
\draw[-Stealth] (we.west) -- ++(-3mm,0) |- (cfg.west);

\node[title, right=2cm of ui] (gc) {graph construction};
\node[elem, below=\hsep of gc] (av) {addVertex};
\node[elem, below=\msep of av] (ap) {addParticle};
\begin{pgfonlayer}{background}
  \node[group, fit=(gc)(av)(ap)] (gcg) {};
\end{pgfonlayer}
\draw[-Stealth] (av.east) -- ++(3mm,0) |- (ap.east);

\node[title, right=2cm of gc] (jt) {\json translation};
\node[elem, below=\hsep of jt] (sv) {saveVertex};
\node[elem, below=\msep of sv] (sp) {saveParticle};
\node[elem, below=\msep of sp] (sc) {saveColorFlow};
\node[elem, below=\msep of sc] (sj) {saveJets};
\begin{pgfonlayer}{background}
  \node[group, fit=(jt)(sv)(sp)(sc)(sj)] (jtg) {};
\end{pgfonlayer}
\draw[-Stealth] (sv.east) -- ++(3mm,0) |- (sp.east);

\node[title, below=0.6cm of gcg] (um) {utilities};
\node[elem, below=\hsep of um] (up) {upload, ...};
\begin{pgfonlayer}{background}
  \node[group, fit=(um)(up)] (umg) {};
\end{pgfonlayer}

\coordinate (m-dat-av) at ($(dat)!0.5!(av)$);
\draw[-Stealth] (dat) -- (m-dat-av |- dat) -- (m-dat-av |- av) -- (av);
\coordinate (m-gcg-jtg) at ($(gcg)!0.5!(jtg)$);
\draw[-Stealth] (dat) -- (m-gcg-jtg |- dat) -- (m-gcg-jtg |- sv) -- (sv);
\draw[-Stealth] (dat) -- (m-gcg-jtg |- dat) -- (m-gcg-jtg |- sc) -- (sc);
\draw[-Stealth] (dat) -- (m-gcg-jtg |- dat) -- (m-gcg-jtg |- sj) -- (sj);
\draw[-Stealth] (umg.west -| uig.east) -- (umg.west);
\draw[-Stealth] (umg.east -| jtg.west) -- (umg.east);

\node[bigtitle, above=\hsep of gcg] (vistas) {\texttt{Vistas}};
\begin{pgfonlayer}{outer}
  \node[outer, fit=(vistas)(uig)(gcg)(jtg)] {};
\end{pgfonlayer}
\end{tikzpicture}
  \end{center}
  \vspace{-1cm}
  \caption{\label{fig:architecture}Code architecture diagram for \vistas.
    White boxes are methods and dark-gray boxes are groupings of methods.
    Arrows indicate method calls between methods.}
\end{figure*}

\vistas is implemented in \python\cite{Python} and consists of two major components:
\begin{enumerate*}[label=(\arabic*)]
\item constructing a directed graph from the \pythia event and
\item translating this graph into a \texttt{JavaScript} Object Notation (\json) structure that can be loaded by the HEP Software Foundation (HSF) browser-based \phoenix event viewer\cite{Moyse:2021xai,Phoenix}.
\end{enumerate*}
A high-level overview of the \vistas code architecture is shown in \cref{fig:architecture} where there is a single class \texttt{Vistas}.
Throughout this section, $\vec{p}_X$ is the three-momentum for particle $X$ in the Lorentz-boosted frame of the visualization.

\subsection{Graph construction}
\label{sec:vistas:graph}

When constructing the directed graph from the \pythia event record there are two important considerations.
First, in \pythia the ``mother-daughter relation normally is reciprocal, but not always''\cite{Pythia}.
This means that traversing the \pythia event record in the forward direction, \eg starting with the first particle in the record and tracing daughter relations, will not always produce the same graph as the backward direction, \eg starting with the final particles and tracing mother relations.
Second, there are a large number of duplicated particle entries which document four-momentum shifts, \eg $\ei{H}{8}$ in \cref{tab:event}, but do not necessarily help with understanding the structure of the event.
For example, in the full event record for the example in \cref{tab:event} there are $633$ particles, but after removing duplicates, this reduces to only $416$.

To address the mother-daughter consideration and ensure a well-defined directed graph, \vistas always traverses the event record backwards, beginning with the last particle.
For visual clarity, \vistas only allows $1 \to N \geq 2$ vertices moving forward in time, except for hard process and MPI vertices.
When a vertex other than a hard process or MPI has multiple mothers in the event record, only the mother that forms the closest angle with the first daughter of the vertex is set as the vertex's mother.
To handle duplicate particles, the graph can be filtered with a selection applied to each particle.
If a particle is not selected, it is not added to the graph and the two vertices linked by the particle are collapsed into a single vertex.

\begin{algorithm} 
  \caption{\label{alg:graph}\vistas graph construction.}
  \begin{algorithmic}[1]
    \State $\vec{v}$ is a dictionary of vertices
    \For{particle $d$ in reversed event}
    \If{$d$ does not pass selection\label{line:select}}
    \State continue
    \EndIf
    \State $v$ is $d$'s production vertex
    \If{$v$ in $\vec{v}$}
    \State add $d$ to $v$'s daughters
    \State set $v$ as $d$'s production vertex
    \Else
    \State create $v$
    \State sort particles $\vec{m}$ by $\cos\theta(\vec{p}_d, \vec{p}_m)$
    \If{$\vec{m}$ from \textit{hard process} or \textit{MPI}}
    \State set $\vec{m}$ as $v$'s mothers
    \State set $v$ as end vertex for $m$ in $\vec{m}$
    \State add $v$ to $\vec{v}$
    \ElsIf{no $m$ in $\vec{m}$ passes selection}
    \State go to line \ref{line:select}
    with first $m$ in $\vec{m}$ as $d$
    \Else
    \State add first selected $m$ to $v$'s mothers
    \State set $v$ as $m$'s end vertex
    \State add $v$ to $\vec{v}$
    \EndIf
    \EndIf
    \EndFor
  \end{algorithmic}
\end{algorithm}

A more concrete description of the \vistas graph construction is given by the pseudo-code in \cref{alg:graph}.
The algorithm builds a dictionary of vertices, $\vec{v} = \{v_i\}$, where for each vertex $v_i$ the algorithm associates a list of daughter particles $\vec{d}_i$ and a list of mother particles $\vec{m}_i$.
This is achieved by first selecting a candidate daughter $d$ and all its mothers $\vec{m}$.
If the mothers $\vec{m}$ are already associated with an existing vertex $v_i$, then the candidate daughter $d$ is appended to the existing list of daughters $\vec{d}_i$ associated with this vertex.
Otherwise, a new vertex $v$ is created and appended to the existing list of vertices, $\vec{v}$.
The list of mothers associated with a given vertex is angular ordered with respect to the first daughter in $\vec{d}$, where the three-momentum of the first mother is closest in angle. 

Note that the default selection applied in \vistas is to only keep particles that have more than one daughter, or are from the hard process or MPI.
This is achieved by the loop body in \cref{alg:graph}, beginning with line \ref{line:select}, which corresponds to the \texttt{addVertex} method of the \texttt{Vistas} class in \cref{fig:architecture}.

The end result of \cref{alg:graph} is a graph structure that is well suited for visualization.
In the constructed graph each particle is either a mother, a daughter, or both.
Particles can now be presented as lines connecting vertices.
A daughter particle has an associated production vertex, where the line begins, and a mother particle has an end vertex, where the line ends.
A daughter particle with no end vertex is a final-state particle, and a mother particle with no production vertex is an initial-state particle, \ie the beams.

\subsection[Translation to JSON structure]{Translation to \json structure}
\label{sec:vistas:json}

In the next step, \vistas translates the above graph into a \json file, which is then used as the input to the \phoenix event viewer.
\Cref{alg:json} details the pseudo-code that creates the \json structure, and in \cref{app:json} example \json from the event of \cref{tab:event} is given.

\vistas uses \texttt{Track}\footnote{
This should not be confused with the HEP-specific slang \textit{track} which is the trajectory of a charged particle measured by detectors such as ATLAS or CMS.
These particle trajectories can be curved, due to the movement of a particle in a detector's magnetic field, or they can be straight lines, if there is no magnetic field.
The \texttt{Track} object in \phoenix retains this capability of representing curved trajectories of particles when visualizing an event.}
objects within the \phoenix framework to display particles.
Each \texttt{Track} object can be assigned a color and the track trajectory is defined by a set of $(x,y,z)$ points with cubic interpolation between them.
The trajectory for particle $X$ is constructed using an appropriately chosen trajectory length-scale factor $l_X$ and the unit vector $\hat{u}_X$ along the direction of particle $X$'s three-momentum, $\hat{u}_X=\vec{p}_X/|\vec{p}_X|$.
By default, $l_X$ is set to a constant value so all particle trajectories have the same length.
However, $l_X$ can also be configured to depend on the properties of particle $X$, \eg $s_x$ could be larger for higher energy particles.
Further details are given in \cref{sec:usage:options}.

\begin{algorithm} 
\caption{\label{alg:json}\vistas translation to \json structure.}
\begin{algorithmic}[1]
  \State $\vec{v}$ is a list of \textit{hard} and \textit{MPI} vertices
  \For{vertex $v$ in $\vec{v}$}
  \State $\vec{x}$ is the vertex position\label{line:position}
  \For{$d$ in $v$'s daughters $\vec{d}$}
  \If{$d$ has no trajectory}\label{line:forward}
  \State $d$'s trajectory is $(\vec{x}, \vec{x} + l_d\hat{u})$
  \If{$d$ has end vertex $e$}
  \State $e$ has position $\vec{x} + l_d\hat{u}_d$
  \State go to line \ref{line:position} with $e$ as $v$
  \EndIf
  \EndIf
  \EndFor
  \For{$m$ in $v$'s mothers $\vec{m}$}\label{line:mothers}
  \If{$m$ has no trajectory}\label{line:backward}
  \State $m$'s trajectory is $(\vec{x}, \vec{x} - l_m\hat{u}_m)$
  \If{$m$ has origin vertex $o$}
  \State $o$ has position $\vec{x} - l_m\hat{u}_m$
  \If{$o$ has daughters besides $m$}
  \State append $o$ to end of $\vec{v}$
  \EndIf
  \State go to line \ref{line:mothers} with $o$ as $v$
  \EndIf
  \EndIf
  \EndFor
  \EndFor
\end{algorithmic}
\end{algorithm}

The position of each hard process and MPI vertex can either be set to the origin of the event display, or alternatively, the hard-process and MPI vertices can be sequentially shifted along the $x$-direction to visually spread them out.
Starting from these hard process and MPI vertices, \vistas then constructs a \json file structure for the remaining vertices, as in \cref{alg:json}.
That is, the \texttt{saveVertex} method of \cref{fig:architecture} first creates the position $\vec{x}$ for each of the hard process and MPI vertices.
In \cref{alg:json} this corresponds to the loop body beginning on line \ref{line:position}.
In the next step, the trajectories of the mothers and daughter particles connected to hard-process and MPI vertices are then constructed by the \texttt{saveParticle} method called by \texttt{saveVertex}.
This is represented by the loop bodies in \cref{alg:json} beginning on line \ref{line:forward} for daughters and \ref{line:backward} for mothers.
Finally, the \texttt{saveVertex} and \texttt{saveParticle} loops are sequentially applied, until all the particle trajectories are constructed. 

A separate approach is required to construct the color-flow lines, shown as dark-gray curved lines in \cref{fig:stages:hadronization}.
Color lines are built by selecting all particles terminating a color flow.
These are particles which carry color/anti-color but have no daughters with color/anti-color.
The color-flow trajectory for a particle with color $c$ and end-vertex position $\vec{x}_c$, and a particle with corresponding anti-color $a = c$ and end-vertex position $\vec{x}_a$, is defined by three points, $(\vec{x}_{c}, \vec{x}_b, \vec{x}_{a})$, where $\vec{x}_b$ is a displaced midpoint that defines the extension of the string arch, and is given by
\begin{equation}
  \label{equ:bow}
  \vec{x}_b = \vec{x}_c + \vec{x}_m + (\vec{x}_c + \vec{x}_a)|\vec{x}_m|/|\vec{x}_c + \vec{x}_a| \mmp{,}
\end{equation}
where
\begin{equation}
  \vec{x}_m = (\vec{x}_a - \vec{x}_c)/2 \mmp{.}
\end{equation}
Quarks and anti-quarks will always be connected to a single color-flow line, while gluons will always be connected to two color-flow lines.
These color-flow trajectories are built with the \texttt{saveColorFlow} method of \cref{fig:architecture}.

Jets, particles clustered using a specified algorithm, can also be optionally included in the visualization and can either be passed by the user or are built with \fastjet\cite{Cacciari:2005hq,Cacciari:2011ma}.
For this, the \texttt{Jet} objects in the \phoenix framework are used, where the radius of the displayed jet cone is the jet radius argument passed to \fastjet, the production vertex is the event origin, and the end vertex is $l_X\vec{p}_X/|\vec{p}_X|$, a scaled unit-vector along the three-momentum of jet $X$.
The jets are built in the \texttt{saveJets} method of \cref{fig:architecture}.

\section[Vistas usage]{\vistas usage}
\label{sec:usage}

Here, we first describe how to install \vistas in \cref{sec:usage:install}, then introduce the user interface in \cref{sec:usage:interface}, followed by how to interact with the event display in \cref{sec:usage:display}, and finally detail more advanced configuration options in \cref{sec:usage:options}.

\subsection{Installation}
\label{sec:usage:install}

Because \vistas is tightly tied to \pythia, it is deployed through the standard \pythia release mechanisms.
This means that \vistas can either be accessed from a user build of \pythia with the \texttt{--with-python} flag passed to the configuration script or through the pre-built binaries for \python via the Python Package Index\cite{PyPi}.

\code{figs/pip.py}

\vistas is designed to work directly through the \python interpreter, as well as within \jupyter\cite{Thomas:2016pnb} or \colab notebooks.
\vistas only uses the built-in \python standard library with no external dependencies other than \pythia.

\subsection{User Interface}
\label{sec:usage:interface}

To use \vistas, a \pythia event must first be generated.
The following generates the event shown in \cref{fig:vistas,fig:stages}.
For details on how to configure \pythia, see the \pythia manual\cite{Pythia}.

\code{figs/pythia.py}

\noindent Next, a \vistas viewer can be created to visualize the event.

\code{figs/vistas.py}

The \texttt{display} method generates the necessary event-data and configuration \json files and then opens a browser window which loads these files into the \phoenix event display.
This is accomplished by temporarily uploading the generated \json files to a publicly accessible URL which is passed via a query string to the \texttt{trackml} instance of the HSF \phoenix event display.
The temporarily hosted \json files must satisfy cross-origin resource sharing (CORS) requirements.
A number of options are available for the \texttt{display} method of the \texttt{Vistas} class; details are available by using \python's built-in \texttt{help} command.
These options include passing modified event data or configurations, where to upload the temporary \json files, and what \phoenix URL to use.

If successful, the \vistas viewer will print a temporary URL to screen, which can be pasted into a browser window if one does not open automatically.

\code[language={}]{figs/url.txt}

\noindent By default, this link is valid for five seconds although the \texttt{sleep} argument to \texttt{display} can be used to control this time window.\footnote{
This \texttt{sleep} is necessary to allow the \phoenix event display to fully load in the browser, before the temporary \json files are deleted.
Ideally, this cleanup would be performed by the server hosting the \json files rather than \vistas.}
The \json can also be written to file with the \texttt{write} command where the \texttt{name} argument specifies the root file names; \texttt{<name>\_data.json} and \texttt{<name>\_cfg.json} will be produced.
These files can then be loaded via the \phoenix graphical user interface.
Abridged examples of these \json files for the event of \cref{tab:event} are given in \cref{app:json}.

\subsection{Display interaction}
\label{sec:usage:display}

\begin{figure}
  \begin{center}
    \includegraphics[width=\columnwidth]{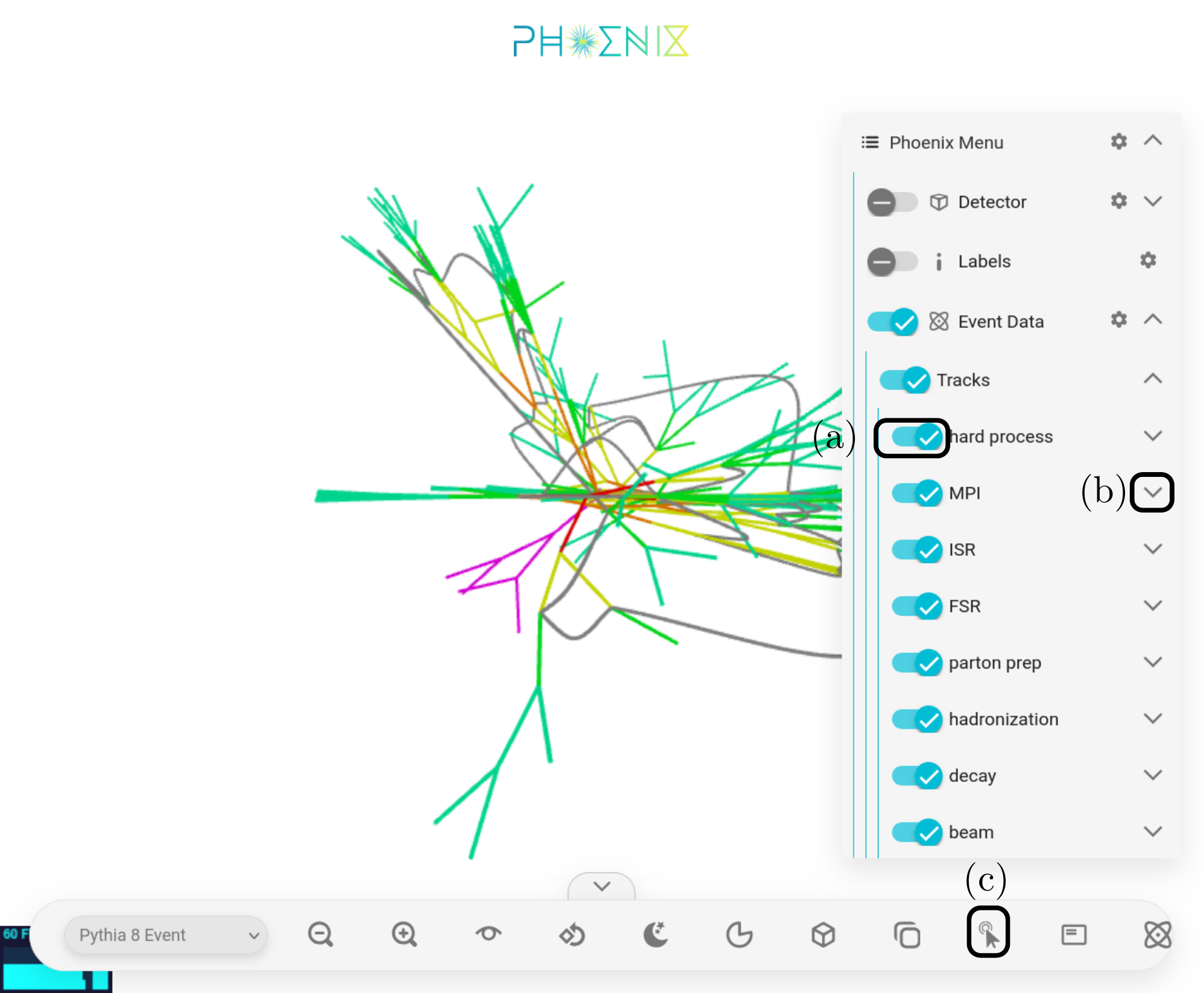}
  \end{center}
  \caption{\label{fig:display}Visualization display interface for the event of \cref{tab:event}.}
\end{figure}

The event display generated by the user interface described in \cref{sec:usage:interface} should look similar to the one shown in \cref{fig:display}.
The event can be rotated by clicking and then dragging on the event.
Hitting the \keys{\texttt{shift}} key, clicking, and then dragging translates the events.
Zooming in and out is possible via standard gestures such as the scroll-wheel of the mouse, or a two fingered swipe on a touch device.

\begin{figure}
  \begin{center}
    \begin{subfigure}{0.46\columnwidth}
      \includegraphics[width=\textwidth]{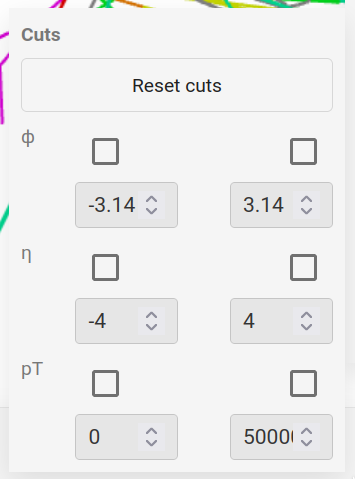}
      \caption{\label{fig:menu:cuts} Particle selection.}
    \end{subfigure}
    \begin{subfigure}{0.46\columnwidth}
      \includegraphics[width=\textwidth]{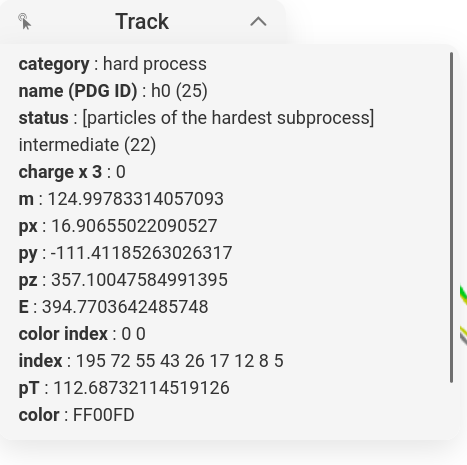}
      \caption{\label{fig:menu:info} Particle information.}
    \end{subfigure}
  \end{center}
  \caption{\label{fig:menu}Example selection requirements and information for the Higgs boson visualized in \cref{fig:display}.}
\end{figure}

The bottom menu bar provides general controls for the event display.
The right menu bar labeled \textit{Phoenix Menu} controls how to display the particles of the event.
Particles from different stages of the event can be turned on or off by clicking the corresponding toggles in the \textit{Phoenix Menu}.
For instance, clicking the toggle in the box labeled (a) in \cref{fig:display} turns off visualization of the hard process.
Clicking the arrow next to each stage toggle, \eg the arrow in box (b) for MPI particles, displays \textit{Draw Options}, \textit{Color Options}, and \textit{Cut Options} for that stage.
With \textit{Cut Options}, selection requirements can be applied to particles in that category for each particle's azimuthal angle ($\phi$), pseudo-rapidity ($\eta$), and transverse momentum ($p_T$), as shown in \cref{fig:menu:cuts}.
Four-momentum related information is always given in the frame the event was generated in, which is not necessarily the same as the visualization frame.
The visualization frame can be selected via the \texttt{frame} option, see \cref{tab:options}, but by default the visualization frame is the rest frame of the first hard process.

Clicking the icon in the black square on the bottom menu, labeled (c) in \cref{fig:display}, allows for individual particles to be selected by hovering over the particle.
Information for that particle, \ie the information discussed in \cref{sec:pythia:event}, is then displayed in the upper left corner of the display as shown in \cref{fig:menu:info}.
The index gives all the indices of the particle, including those not displayed due to collapsing vertices in the event.
In the example of \cref{fig:menu:info}, $\ei{H}{195}$ is the final instance of the Higgs boson in the event record of \cref{tab:event}, before decaying into $\ei{Z}{257}\, \ei{Z}{258}$, while $\ei{H}{5}$ is the first instance generated from the hard process.
These indices can be useful for tracing a particle through the \pythia event record.

\phoenix also provides a virtual reality (VR) mode that can be activated, assuming appropriate hardware, by clicking the goggles icon in the bottom menu bar; this is not visible in \cref{fig:display}.
\vistas had been  already used to great effect in VR mode with a MetaQuest 3 at various summer schools, workshops, and conferences.

\subsection{Advanced Options}
\label{sec:usage:options}

\begin{table*}
  \caption{\label{tab:options}Options for event visualization.
    Up-to-date options can always be found by calling \python's built-in \texttt{help} on the \texttt{option} method of a \texttt{Vistas} object.}
  \begin{tabular}{@{}p{0.20\textwidth} p{0.13\textwidth} p{0.60\textwidth}@{}}
  \toprule
  \textbf{option} & \textbf{type} & \textbf{description} \\
  \midrule
  
  \texttt{frame} & \texttt{str} & boost the event into a specific frame: \texttt{None} (no boost), \texttt{"hard process"} (boost to the hard-process frame), \texttt{"beam"} (boost to the beam frame). \\
  \midrule
  
  \texttt{show} & \texttt{[str, \dots]} & categories toggled on for display. Toggled-off categories can still be re-enabled. \\
  \midrule
  
  \texttt{highlight} & \texttt{[str, \dots]} & categories to highlight; all others are grayed out. An empty list disables graying. \\
  \midrule
  
  \texttt{mpi} & \texttt{float} & sub-collision separation factor; multiplied by \texttt{length:factor} to set the separation. \\
  \midrule
  
  \texttt{jets} & \texttt{dict} & controls jet building and display (maps to \pythia \texttt{SlowJet} arguments). \\
  \quad \texttt{algorithm} & \texttt{str} & jet building algorithm: \texttt{None} (no jets), \texttt{"akt"} (anti-$k_T$), \texttt{"ca"} (Cambridge/Aachen), \texttt{"kt"} ($k_T$). \\
  \quad \texttt{r} & \texttt{float} & jet size parameter (jet cone radius). \\
  \quad \texttt{ptmin} & \texttt{float} & minimum transverse momentum of the jets. \\
  \quad \texttt{etamax} & \texttt{float} & maximum pseudorapidity of particles used in jet building. Values above 20 disable the cut. \\
  \quad \texttt{select} & \texttt{str} & particles used: \texttt{"all"}, \texttt{"visible"} (no neutrinos), or \texttt{"charged"}. \\
  \quad \texttt{mass} & \texttt{str} & particle mass assignment: \texttt{"zero"} (massless), \texttt{"pion"} (pion mass for non-photons), \texttt{"gen"} (generated mass). \\
  \quad \texttt{length} & \texttt{dict} & drawing options for jets; same fields as the \texttt{length} dictionary below. \\
  \midrule
  
  \texttt{length}, \texttt{color} & \texttt{dict} & control length and color attributes of particle lines. \\
  \quad \texttt{scale} & \texttt{str} & \texttt{"constant"} (fixed) or \texttt{"log"} (logarithmic with variable skew). \\
  \quad \texttt{observable} & \texttt{str} & observable used for the log scale; any valid \python code where \texttt{p} is the \pythia \texttt{Particle} may be passed; \texttt{"p.e()"} and \texttt{"p.pT()"} would use the particle's energy or transverse momentum, respectively. For jets, \texttt{p} is a \pythia \texttt{Vec4} object rather than a \texttt{Particle} object. Kinematic methods between these two object types are identical. \\
  \quad \texttt{skew} & \texttt{float} & must satisfy $|\text{\texttt{skew}}| > 1$; values of $\pm 1$ give a linear scale. Larger positive/negative skew stretches smaller/larger values more. \\
  \quad \texttt{group} & \texttt{str} & \texttt{"all"} (use all categories) or \texttt{"cat"} (use the particle's own category). \\
  \quad \texttt{factor} & \texttt{float} & (\texttt{length} only) multiply the scale by this factor. \\
  \quad \texttt{offset} & \texttt{float} & (\texttt{length} only) add this value to the scale. \\
  \quad \texttt{min} / \texttt{max} & \texttt{float} & (\texttt{color} only) shading limits ($-1$ white/low, $+1$ black/high). If \texttt{max} $<$ \texttt{min}, the mapping inverts. \\
  \midrule
  
  \texttt{verbosity} & \texttt{dict} & logging levels: integer keys, string values. Removing an entry suppresses that level. \\
  \bottomrule
\end{tabular}

\end{table*}

\begin{figure*}
  \begin{center}
    \begin{subfigure}{0.47\textwidth}
      \display{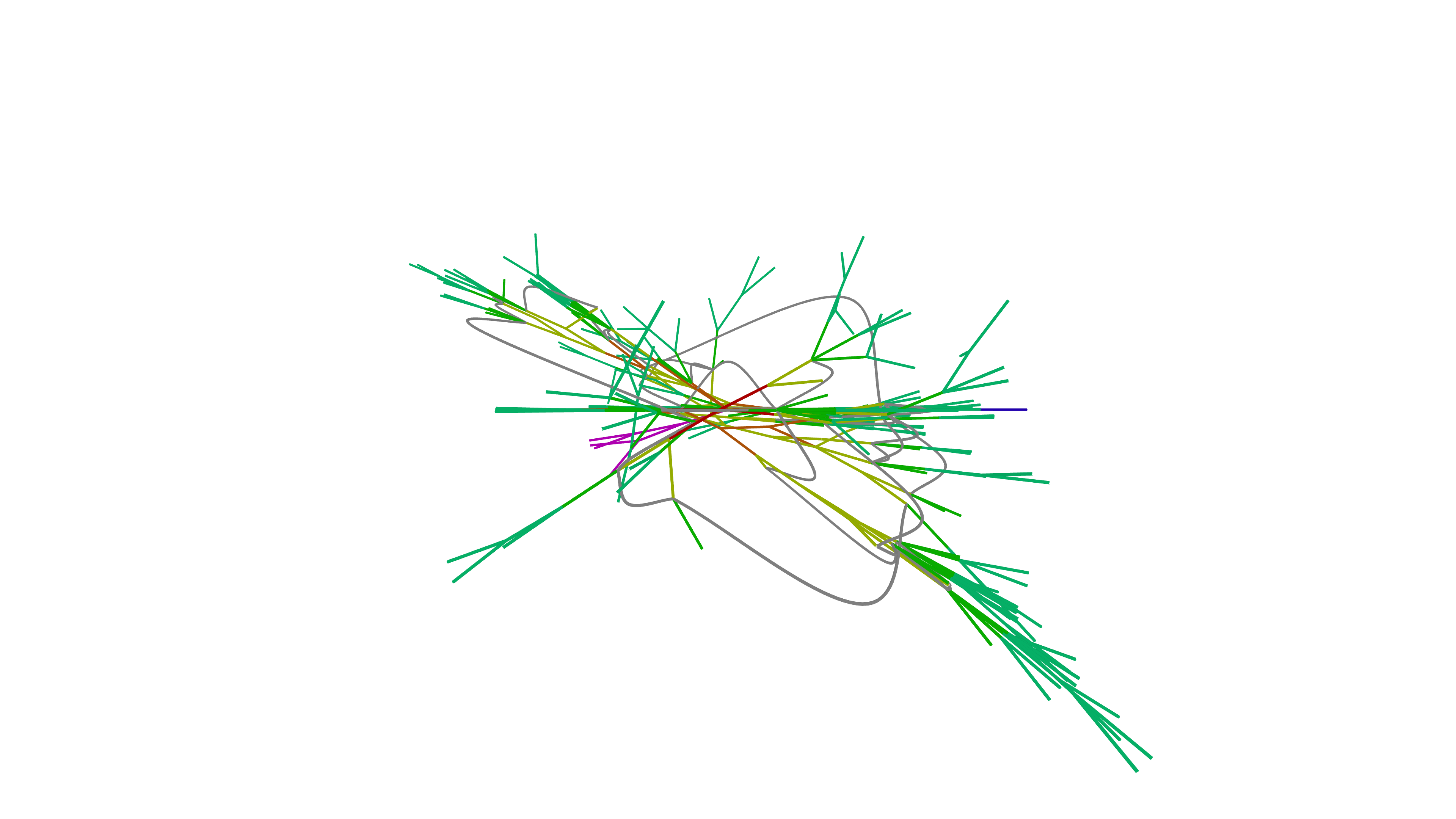}
      \caption{\label{fig:options:boost} Lab frame visualization.}
    \end{subfigure} \hfill
    \begin{subfigure}{0.47\textwidth}
      \display{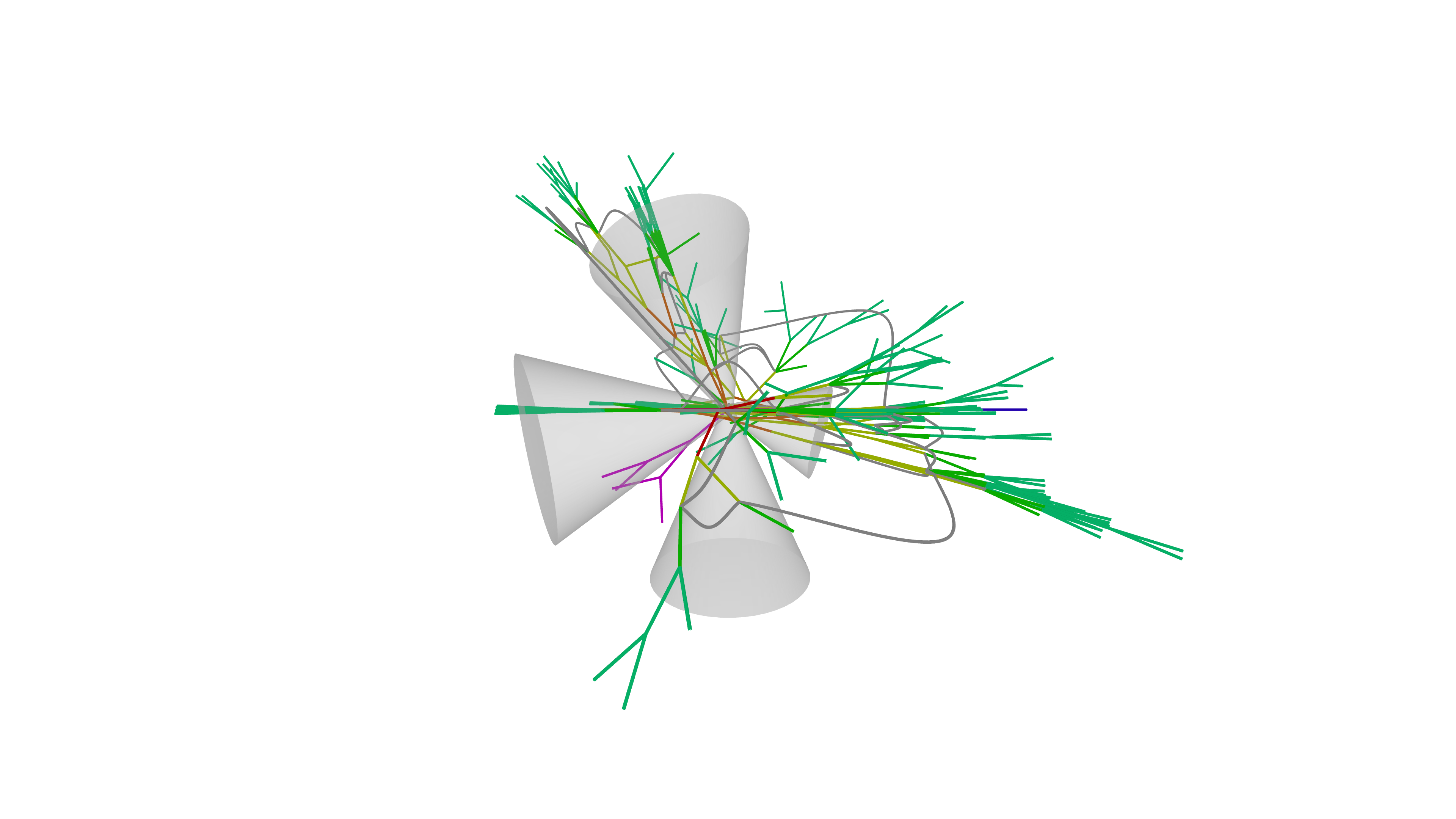}
      \caption{\label{fig:options:jets} Jet visualization.}
    \end{subfigure} \\
    \begin{subfigure}{0.47\textwidth}
      \display{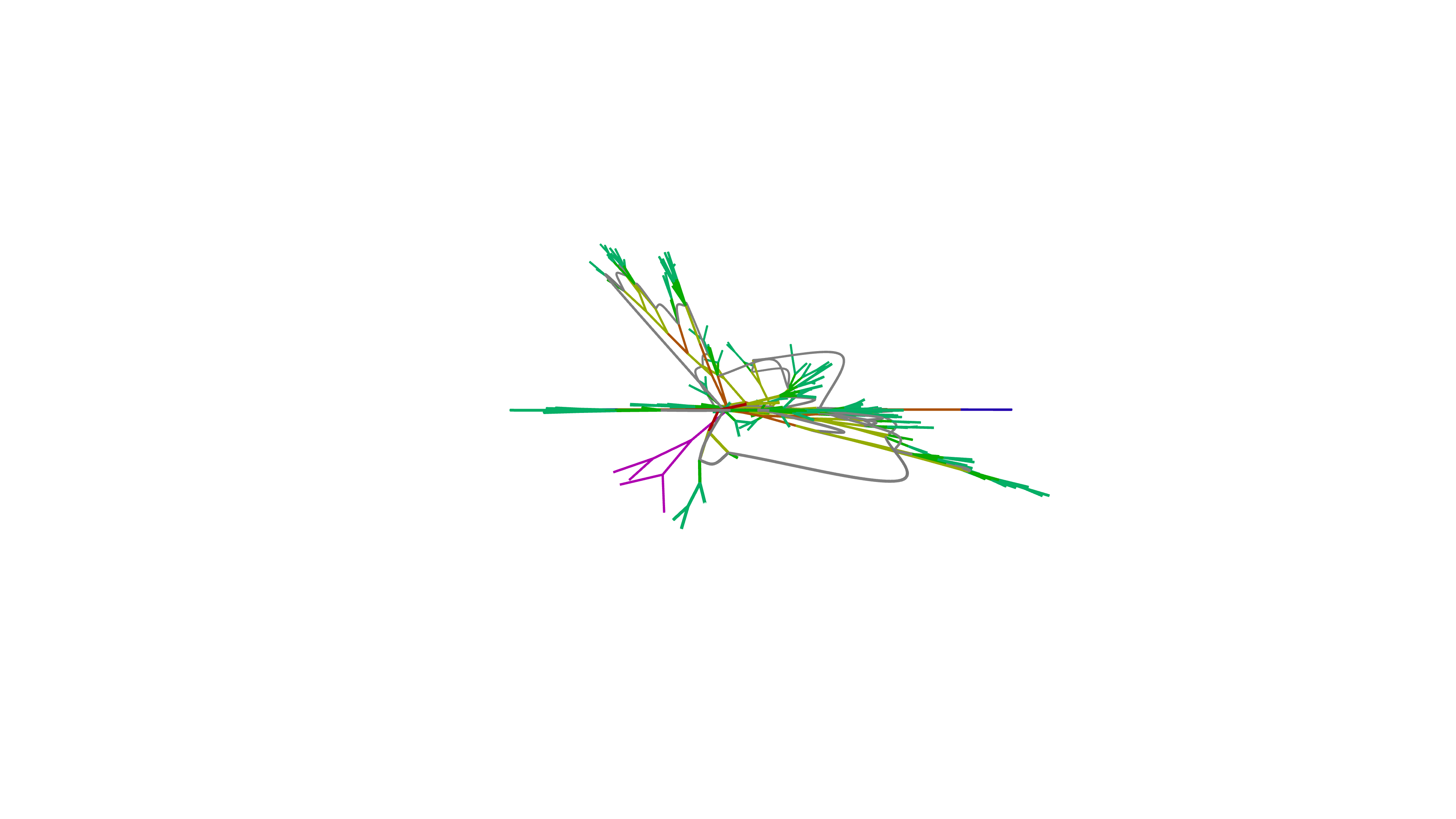}
      \caption{\label{fig:options:length} Logarithmic length scale in particle energy.}
    \end{subfigure} \hfill
    \begin{subfigure}{0.47\textwidth}
      \display{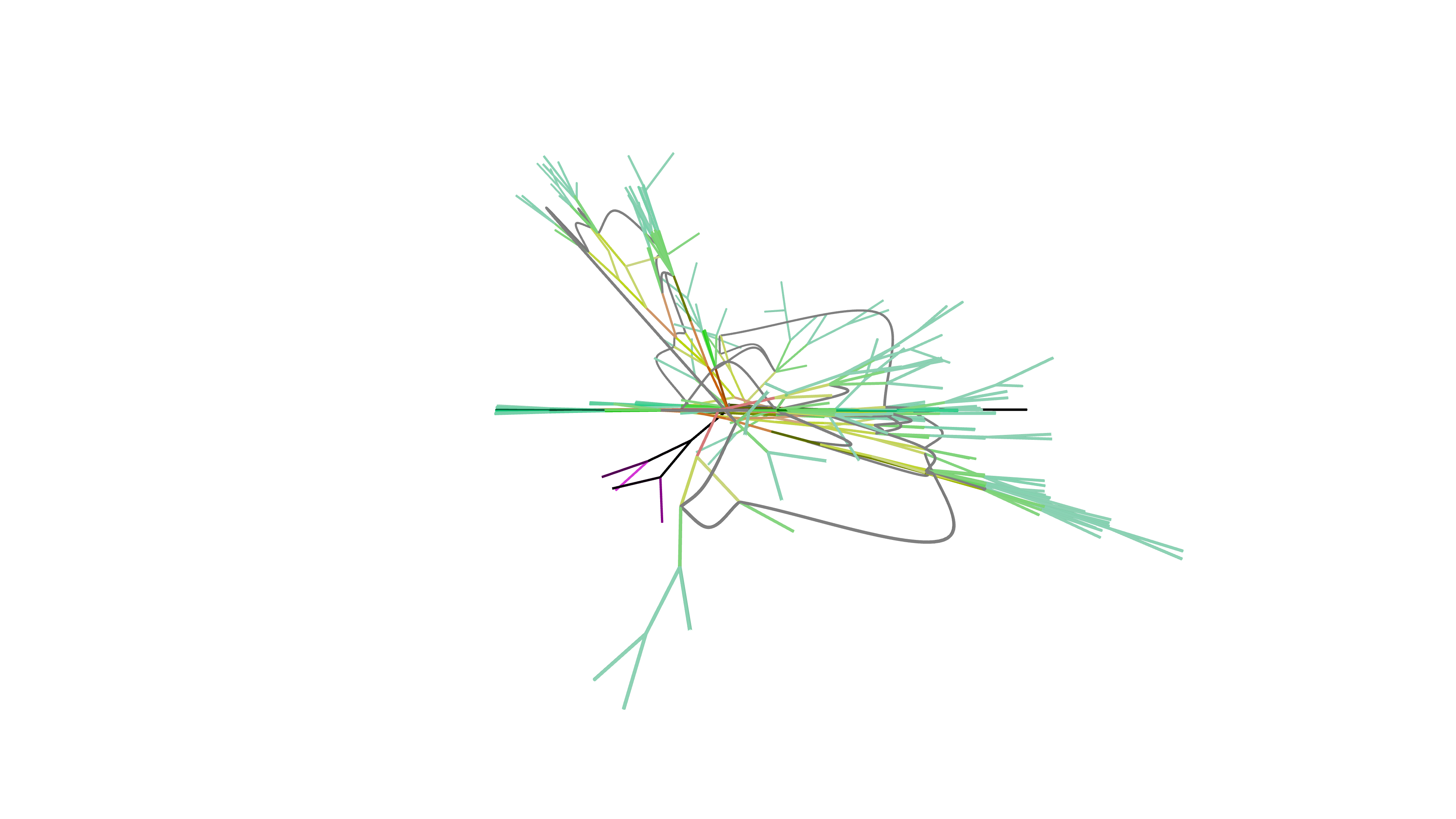}
      \caption{\label{fig:options:color} Logarithmic color scale in particle energy.}
    \end{subfigure}
  \end{center}
  \caption{\label{fig:options}Example visualization options available in \vistas, as defined in \cref{tab:options}.
    In all four images, the event is the same Higgs event as in \cref{tab:event}.}
\end{figure*}

A full list of available options in \vistas, at the time of writing, is given in \cref{tab:options}.
Since \vistas will be updated over time, using \python's built-in \texttt{help} method on the \texttt{option} method of a \texttt{Vistas} object should always return the most up-to-date list of options available for that version of \vistas.
These options can be set via the \texttt{opts} dictionary of a \texttt{Vistas} object.

\code{figs/options.py}

As shown above, the default options can always be recovered by setting the \texttt{opts} dictionary to the output of the \texttt{options} method.
In \cref{fig:options} a few of the available options are demonstrated.
\Cref{fig:options:boost} shows \cref{fig:vistas} but in the laboratory frame rather than the hard process frame and \cref{fig:options:jets} shows the option of turning on jet building.

The trajectory length or color for particles can be set proportional to a particle observable $\mathcal{O}$, \eg the particle's energy, rather than a fixed scale.
Changing the trajectory length is achieved by modifying for each particle the value of the length-scale factor $l_X$ introduced in \cref{sec:vistas:json}.
First the relative value of the observable is calculated
\begin{equation}
  r_X = (\mathcal{O}_X - \mathcal{O}_{\min})/(\mathcal{O}_{\max} -
  \mathcal{O}_{\max}) \mmp{,}
\end{equation}
where $\mathcal{O}_{\max/\min}$ are the maximum/minimum values of the observable for all particles in that stage of the \pythia simulation.
Alternatively, the minimum and maximum values of the observable for all particles visualized in the event can be used by setting the \texttt{group} option to \texttt{"all"}.
The observable is set by the \texttt{observable} option, see \cref{tab:options}, and can be any valid \python code where the particle is referenced as \texttt{p} and is of the type \texttt{Particle} from \pythia.
For example, \texttt{math.log(p.e())} would set $\mathcal{O}_X$ to the logarithm of the particle's energy while \texttt{p.tau()} would set $\mathcal{O}_X$ to the particle's invariant lifetime.
For jets, \texttt{p} is of the type \texttt{Vec4} from \pythia, which has the same kinematic methods as \texttt{Particle}, \eg \texttt{e} or \texttt{pT}.

For each particle $X$ the value $r_X$ is then mapped onto a logarithmic scale,
\begin{equation}
  t_X =
  \begin{cases}
    \log_k(1 + (k - 1)r_X) & k > 1 \\
    1 - \log_k(1 + (k - 1)(1 - r_X)) & k < -1 \\
  \end{cases}
  \mmp{,}
\end{equation}
where $k$ is the skew of the mapping set by the \texttt{skew} option.
A large positive $k$ spreads out small values of $r_X$ while a large negative value of $k$ spreads out large values of $r_X$.
As $k$ approaches $\pm 1$ the rescaling becomes linear.

The length-scale factor $l_X$ is given by,
\begin{equation}
  l_X = d_\text{\texttt{fac}} t_X + d_\text{\texttt{off}} \mmp{,}
\end{equation}
where $d_\text{\texttt{fac}}$ is set by \texttt{factor} and $d_\text{\texttt{off}}$ by \texttt{offset}.
The color scale is set by a similar factor,
\begin{equation}
  c_X = (d_\text{\texttt{max}} - d_\text{\texttt{min}}) t_X + d_\text{\texttt{min}} \mmp{,}
\end{equation}
where $d_\text{\texttt{max}/\texttt{min}}$ are set by \texttt{max} and \texttt{min} and must fall between $-1$ (white) and $+1$ (black) and a value of $0$ corresponds to the unmodified base color for each simulation stage.

An example of the logarithmic scale based on particle energy is shown in \cref{fig:options:length} for length and \cref{fig:options:color} for color.

\section{Conclusions}
\label{sec:conclusions}

\begin{figure*}
  \begin{center}
    \includegraphics[width=\textwidth]{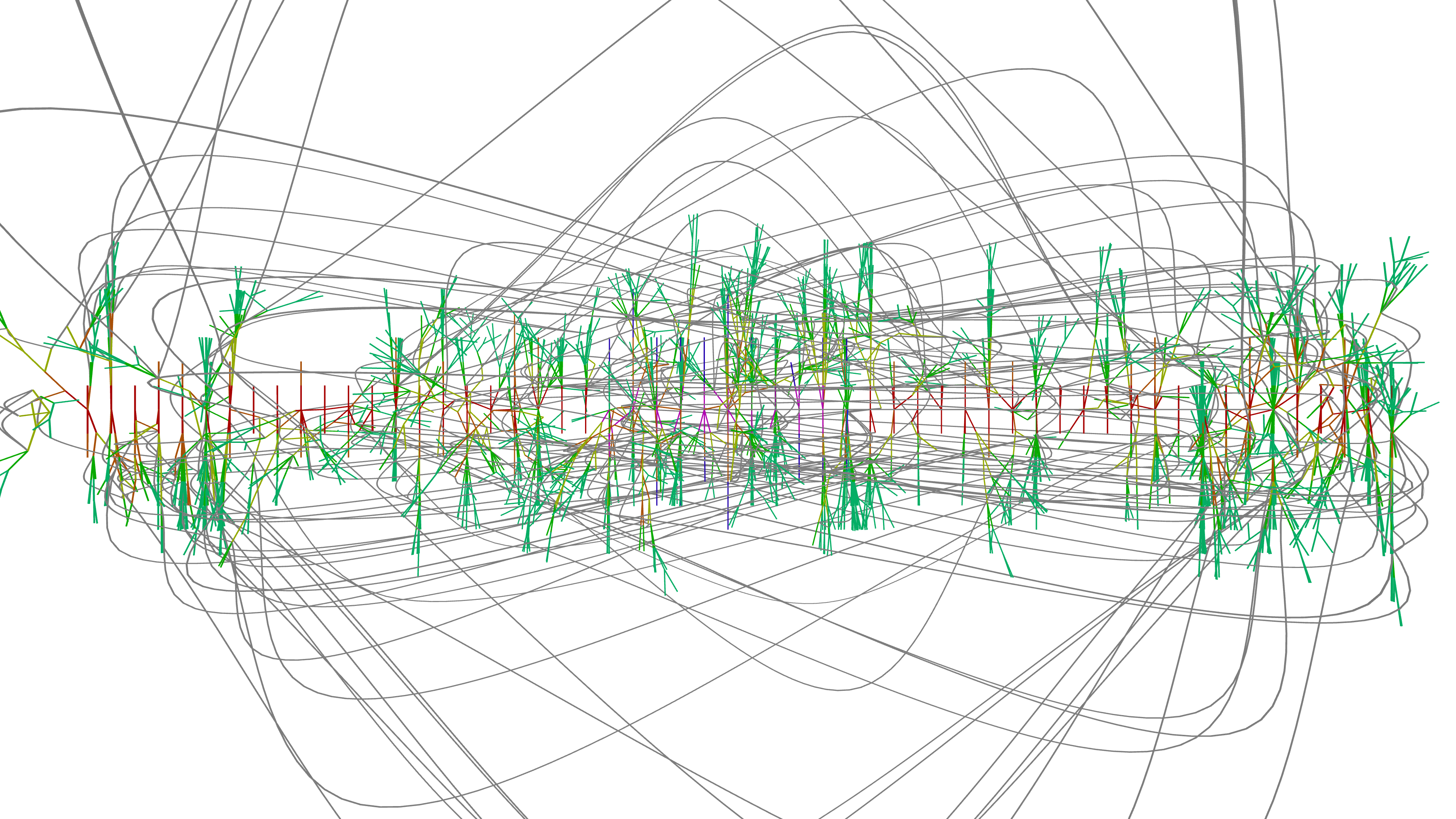}
  \end{center}
  \caption{\label{fig:hi}Example heavy ion event visualized with \vistas.}
\end{figure*}

In this paper we introduced a visualization tool, \vistas, for high-energy particle collision simulations.
This new visualization tool is distinct from previous tools for visualizing high energy events, which focus on visualizing reconstructed objects in particle physics detectors.
\vistas instead focuses on the underlying particle physics process, visualizing the \pythia event record.
\vistas organizes the particles generated in the simulation into several sets, based on which algorithmic stage of the \pythia simulation they were generated in: beams, hard process and multi-parton interactions, initial- and final-state radiation, hadronization, and decays.
The color flow of QCD charge is also visualized, and optionally, jets of particles can be displayed.
\vistas generates an interactive three-dimensional display of an event, using the HSF \phoenix browser-based interface.
The user can access the full utility of \phoenix: zoom, translate, display information on particles, select specific stages of the event, and apply kinematic requirements to particles.
This organization and interactive environment allows for an intuitive understanding of the underlying physics to be developed for simulated events.

The underlying architecture of \vistas is designed to be as general and robust as possible.
Consequently, \vistas can in principle be used for all \pythia physics processes, although this has not been exhaustively validated.
Such an example stress test of \vistas is shown for a heavy-ion event in \cref{fig:hi} with a particle multiplicity of $6,072$.

Due to its interactive nature, we anticipate that \vistas can be useful in a number of educational settings.
For instance, we found that the virtual reality mode of the \phoenix browser-based interface is a popular feature when a developmental version of \vistas was used in outreach efforts.
Fundamental physics laws can be easily seen through \vistas, such as conservation of momentum and energy, or QCD charge.
Characteristic features of event types can also be quickly identified.
For example, the soft and collinear structure of QCD is readily visible, or the rapidity gap due to color flow in processes like vector-boson fusion production of the Higgs boson.
Depending on learning objectives, \vistas can be easily configured to intuitively demonstrate a wide range of phenomena in high-energy particle physics.

\section*{Acknowledgments}
We thank Edward Moyse, Ben Courtier, and Sebastien Ponce for their support in working with \phoenix.
We also thank Pavel Nadolsky, Fred Olness, and the CTEQ summer school for providing a platform for beta testing \vistas, as well as Alexei Prokudin and the CFNS summer school.
Finally, we thank the \pythia collaboration for their support of \mlhad work, particularly Peter Skands and Torbj\"orn Sj\"ostrand for their comments on this manuscript.

BA, JZ, and RG acknowledge support in part by the DOE grants DE-SC0011784 and DE-SC0026301, and NSF  grants OAC-2103889, OAC-2411215, and OAC-2417682.
JZ, TM, and MW also acknowledge support in part from the Visiting Scholars Award Program of the Universities Research Association.
PI and MW are supported by NSF grants OAC-2103889, OAC-2411215, OAC-2417682, and NSF-PHY-2209769.
CB acknowledges support from Vetenskapsrådet contract number 2023-04316.

This document was prepared using the resources of the Fermi National Accelerator Laboratory (Fermilab), a U.S. Department of Energy, Office of Science, Office of High Energy Physics HEP User Facility.
Fermilab is managed by FermiForward Discovery Group, LLC, acting under Contract No. 89243024CSC000002.

\bibliographystyle{utphys}
\bibliography{main}

\appendix
\section{Code}
\label{app:code}

The \vistas code and an associated tutorial are available at the following repositories.
\begin{itemize}
\item The code for \vistas used in this paper is available at: \\
\url{https://gitlab.com/Pythia8/releases/-/blob/pythia8318/plugins/python/extra/Vistas.py}
\item Future versions of \vistas will be available at: \\
 \url{https://gitlab.com/Pythia8/releases/-/blob/master/plugins/python/extra/Vistas.py}
 \item An interactive \jupyter notebook that explores \pythia using \vistas is available through the \textit{M}onte \textit{C}arlo \textit{G}eneral \textit{E}ducation \textit{N}etwork (\textsc{MCgen}) at: \\
  \url{https://gitlab.com/mcgen-ct/tutorials/-/blob/main/.solutions/pythia/vistas.ipynb}
\item The above notebook was provided for the 2026 \textit{C}enter for \textit{F}rontiers in \textit{N}uclear \textit{S}cience (CFNS) via \colab at: \\
   \url{https://colab.research.google.com/github/mcgen-ct/tutorials/blob/2026-cfns/.solutions/pythia/vistas.ipynb} 
\end{itemize}

The tutorial steps through the stages of an MCEG, beginning with an $e^+\, e^- \to Z \to \mu^+\, \mu^-$ simulation of the Large Electron Positron (LEP) collider.
Rather than toggling on or off parts of the visualization, the notebook directly configures \pythia to only run certain stages of the simulation.

First, only the the hard process is turned on with \pythia, followed by adding ISR and then FSR.
Next, decays are demonstrated by switching the $Z$ decay to $Z \to \tau^+\, \tau^-$.
Finally, hadronization is shown by using the decay $Z \to u, \bar{u}$.
Further examples are given with an Electron Ion Collider (EIC) configuration of $e^-\, p \to X$, LHC configurations $p\, p \to t\, \bar{t}$ and $q\, \bar{q} \to q\, \bar{q}\, H$, and an LHC heavy-ion configuration $\text{Pb}\, \text{Pb} \to X$.
Exercises are also given exploring visualization options.
The two notebook links given above are for the solution versions of these notebooks.
Exercise versions of these notebooks are available by removing \texttt{.solutions/} from the URL.

An earlier version of \vistas was used for the 2025 \textit{C}oordinated \textit{T}heoretical-\textit{E}xperimental Project on \textit{Q}CD (CTEQ) summer school, with notebooks and code available at \url{https://gitlab.com/mcgen-ct/tutorials/-/tree/2025-cteq/vistas}.
There are a number of differences between this prototype version and the version presented here: automatic uploading was not available, toggling of stages of the event was not possible, and the parsing of the \pythia event was more fragile.

The code used to produce the figures of this paper is given below using \pythia \texttt{8.317}.
The default colors of \vistas are modified slightly in the \texttt{default} method to colors more suitable for printing.
Additionally, the default color dependence on energy is turned off.
The \texttt{capture} method allows the event listing of \pythia to be captured from standard \cpp output and writes this output to an event text file, while the \texttt{display} method allows the camera position to be set by modifying the event configuration \json generated by \vistas.

\code{code/figures.py}

\section[Example JSON]{Example \json}
\label{app:json}

The following \json snippets provide examples of the event data output of \vistas.
An example \json for a particle line is given in \cref{tab:json:hard}.
This is the hard process Higgs boson, $\ei{H}{5}$, for the event of \cref{tab:event} and \cref{fig:vistas}.
The \texttt{"pos"} list corresponds to the positions $\vec{x}$ which define the particle trajectory.
Note that rather than two positions, three positions are defined.
This is because with only two positions, some versions of \phoenix may try to render a non-linear interpolation of the line.
Here, the first position is $\vec{x}$, the third position is $\vec{x} + l_d\hat{u}$, and the second position is the average of the two.
The \texttt{saveParticle} method of the \texttt{Vistas} class produced this \json.

\begin{table}
  \caption{\label{tab:json:hard}Example of the \json event data for the hard-process Higgs boson, $\ei{H}{5}$, from \cref{tab:event} and \cref{fig:vistas}.}
  \code[linerange={2-50}][nbbox]{figs/vistas_data.json}
\end{table}

An example color-flow line is given in \cref{tab:json:color}.
This corresponds to the first color connection of the string example of \cref{tab:event} and \cref{fig:vistas} from \cref{sec:pythia:stages}.
This connects the quark $\ei{u}{275}$ with color \texttt{106} to the gluon $\ei{g}{276}$ with anti-color \texttt{106} and color \texttt{149}.
Three positions in the \texttt{"pos"} list define this line, with cubic interpolation between.
This list is $(\vec{x}_c, \vec{x}_b, \vec{x}_a)$ where $\vec{x}_c$ is the end position of the particle providing the color and $\vec{x}_a$ is the end position of the particle providing the anti-color.
The middle point is defined by \cref{equ:bow}.
The \texttt{saveColorFlow} method of the \texttt{Vistas} class produced this \json.

\begin{table}
  \caption{\label{tab:json:color}Example of the \json event data for the color flow $\ei{u}{275}$ to $\ei{g}{276}$ from \cref{tab:event} and \cref{fig:vistas}.}
  \code[linerange={16865-16884}][nbbox]{figs/vistas_data.json}
\end{table}

An example event configuration \json snippet is given in \cref{tab:json:cfg} for \cref{fig:vistas}.
This configuration does not significantly change between events except for the following:
\begin{enumerate*}[label=(\arabic*)]
\item the \texttt{show} option which allows the user to toggle on or off parts of the event;
\item the definition of the event stages which is controlled by the category database \texttt{cdb}; and
\item the camera position for the event which depends on the topology of the event and is set so the full event should be in view without user adjustment.
\end{enumerate*}
In the default configuration of the \texttt{trackml} instance of \phoenix, particle selection requirements are switched on and can lead to missing particles in the initial view if not explicitly switched off.
This configuration switches off all particle selection requirements and detector elements.

\begin{table}
  \caption{\label{tab:json:cfg}Example of the \json event configuration for the hard process toggle for \cref{fig:vistas}.}
  \code[linerange={67-86}][nbbox]{figs/vistas_cfg.json}
\end{table}

\end{document}